\font\bigtenrm=cmr10 scaled\magstep5
\line{\hfill First Draft Begun September 11, 2001}
\bigskip
\bigskip
\bigskip
\bigskip
\centerline{\bf{\bigtenrm Genesis:}}
 \bigskip
\bigskip
\centerline{\bf{\bigtenrm How the Universe Began}}
\bigskip
\bigskip
\centerline{\bf{\bigtenrm According to}}
\bigskip
\bigskip
\centerline{\bf{\bigtenrm Standard Model Particle
Physics}}
\bigskip
\bigskip

\centerline {by}

\medskip
\centerline {Frank J. Tipler\footnote{$^1$}{e-mail
address: TIPLER@TULANE.EDU}}
\centerline {Department of Mathematics and
Department of Physics}
\centerline {Tulane University}
\centerline {New Orleans, Louisiana 70118 USA}
\bigskip

\vfill\eject
\centerline{{\bf Abstract}}
\bigskip

\noindent
I show that the mutual consistency of the Bekenstein
Bound, the Standard  Model (SM) of particle physics, and
general relativity implies that the universe began in a
unique state, an initial Friedmann $S^3$ singularity at
which the temperature, entropy, Higgs field, baryon
number, and lepton number were zero, but with a
non-zero SU(2) (gravitational) sphaleron field.  I solve
the coupled EYM equations for this unique state, show
how the horizon problem is solved, and how SM
baryogenesis naturally results from the triangle
anomoly.  Since the SU(2) winding number state is thus
non-zero, the universe is not in the QCD ground
state, and this plausibly yields a (small) positive
cosmological constant.  Since the initial state is
unique, it is necessarily homogeneous and isotropic, as
required by the Bekenstein Bound.  Wheeler-DeWitt
quantization implies an $S^3$ cosmology must be very
close to flat if the universe is to be classical today.  I
show that the spectrum of any classical gauge field (or
interacting massless scalar field) in a FRW universe
necessarily obeys the Wien displacement law and the
corresponding quantized field the Planck distribution
law with the reciprocal of the scale factor
$R$ playing the role of temperature, even if the fields
have zero temperature.  Thus the CBR could even today
be a pure SU(2) electroweak field at zero temperature
coupled to the Higgs field, in spite of early universe
inverse double Compton and thermal bremsstrahlung.  I
conjecture that such a pair of fields with this coupling
can yield a weakly inteacting component with mass
density decreasing as $R^{-3}$ and an EM interacting
component with mass density decreasing as $R^{-4}$,
the former being the dark matter and the latter the
CBR.  Except that such a CBR would not couple to
right-handed electrons, and this property can be
detected with a Penning trap or even using  the late
1960's CBR detector with appropriate filters.  I argue
that right handed ultrahigh energy cosmic ray protons
would not produce pions by interacting with such a
CBR, and thus the existence of such protons may
constitute an observation of this CBR property.   I show
that such a CBR has no effect on early universe
nucleosynthesis, and no effect on the location of the
acoustic peaks.

\vfill\eject
\centerline{\bf TABLE OF CONTENTS}

\bigskip
\noindent
1. Introduction

\smallskip
\noindent
2. Apparent Inconsistences in the Physical Laws in
the Early Universe

\smallskip
a. Bekenstein Bound Inconsistent with Second Law
of Thermodynamics
\smallskip
b. Universe NOT Planck-sized at Planck Time

\smallskip
c. FRW Universe does NOT admit a $U(1)$ gauge
field like electromagnetism

\medskip
\noindent
3. The Spectra of Gauge Fields in FRW Background

\smallskip
a. Proof that ALL classical gauge fields necessarily
obey Wein Displacement Law
\settabs 9 \columns
\+&in a FRW universe\cr

\smallskip
b. Proof that ALL quantized gauge fields necessarily
have a Planckian spectrum 
\+&in a FRW universe\cr

\medskip
\noindent
4. Particle Production Solution to EYM equation in FRW
Universe 

\smallskip
a. Exact Solution of EYM Equations with Constant SU(2)
Curvature

\medskip
\noindent
5. Particle Production by Instanton Tunnelling in FRW
Universe

\medskip
\noindent
6. The Unique Quantized FRW Universe
\smallskip
a. Conformal Time is the Unique Physical Time 

\smallskip
b. Consistency between Copenhagen and Many-Worlds
Interpretations Requires
\+& a delta function initial Boundary Condition\cr

\smallskip
c. Solution to Flatness Problem in Cosmology

\smallskip
d. Solution to the Homogeneity, Isotropy, and Horizon
Cosmological Problems

\medskip
\noindent
7. The SU(2) gauge field and the Higgs Field in the
present day epoch

\smallskip
a. splitting of two fields into ``matter'' and ``radiation''
fields

\smallskip
b. Solution to ``dark matter'' problem: what it is and
how it has eluded detection

\smallskip
c. Why an SU(2) component would have no effect on
early universe nucleosynthesis

\smallskip
d. Suppressing early universe pair creation, IC, and
thermal Bremsstrahlung

\medskip
\noindent
8.  Detecting an SU(2) component to the CBR

\smallskip
a. Right-handed electrons Won't Couple to an SU(2)
component in the CBR

\smallskip
b.  Detecting an SU(2) component to the CBR Using Hans
Dehmelt's Penning Trap

\smallskip
 c. Detecting an SU(2) Component With the Original
CMBR Detectors with Filters

\smallskip
d. d. Other Means of Detecting an SU(2) CMBR
Component 

\medskip
\noindent
9.  Has an SU(2) CBR component already been detected?

\smallskip
a. The ultrahigh energy cosmic ray spectrum

\smallskip
b. Why the ultrahigh energy cosmic rays should not
exist, yet they do

\smallskip
c. How an SU(2) component to the CBR would permit the
cosmic rays to propagate.

\medskip
\noindent 10. Conclusion: Advantages of Theory

\smallskip
a. Uses only physics that has actually been seen in
laboratory

\smallskip
b. Solves Flatness, Homogeneity, and Flatness Problems
using only known physics

\smallskip
c. explains why there is more matter than antimatter
in universe

\smallskip
d. Solves Dark Matter Problem using only known physics

\smallskip
e. Solves Cosmological Constant Problem

\smallskip
f. Explains acceleration of universe in current epoch

\+& 1. acceleration due to effective cosmological
constant \cr

\+& 2. gives mechanism that will eventually cancel
accelertion \cr

\smallskip
g. Explains how it is possible for ultrahigh energy
cosmic rays to exist

\vfill\eject

\centerline{\bf{\bigtenrm 1. Introduction}}
\bigskip
\bigskip
It is generally agreed that the non-zero baryon number
of the universe requires explanation.  Most
baryogenesis scenarios envisage some baryon number
violating process occurring at high temperature,
typically at the GUT scale or at the Planck
temperature of $10^{19}$ GeV.  But a high temperature
implies a high entropy density $\sigma$, since
$\sigma \propto T^3$, and thus these scenarios leave
open the question of where the entropy came from.  For
the most natural initial value of the universal entropy
is zero, just as the most natural initial value of the
baryon number is zero.

\medskip
I shall show that the physical laws impose their own
unique boundary condition on the universe, requiring
the universe to begin in a unique state of zero
entropy, zero temperature, zero baryon number and zero
lepton number.  Such a universe must be topologically
$S^3$, initially perfectly isotropic and homogeneous,
with zero Higgs field, but with a non-zero SU(2)
sphaleron field.  I shall give the exact solution to the
Einstein-Yang-Mills equations for this unique initial
state, and show how it evolves.  I shall propose two
experiments to test my model for the early universe.

\medskip
{\bf Acknowledgements:}  This work was supported in
part by the Georges Lurcy Research Fund, and by the
Tulane University Physics Department.  I am grateful
for helpful discussions with James Bryan, David
Deutsch, Paul Frampton, Alan Goodman, Gordon Kane,
John Moffet, Don Page, Bruce Partridge, John Perdew,
George Rosensteel, Simon White, and David Wilkinson.

\vfill\eject

\centerline{\bf{\bigtenrm 2. Apparent Inconsistences
in the }}
 \bigskip
\centerline{\bf{\bigtenrm Physical Laws in the Early
Universe}}

\bigskip
\bigskip
{\bf a. Bekenstein Bound Inconsistent with Second Law
of Thermodynamics}
\bigskip

The fundamental limitation on the number of possible
quantum states in a bounded region --- or,
alternatively,  on the number of bits that can be coded
in a bounded region --- is given by the Bekenstein
Bound [1,2].  The Bekenstein Bound is a consequence of
the basic postulates of quantum field theory.  A
derivation will not be given here, but in essentials
the Bekenstein Bound is just another manifestation
of the Heisenberg Uncertainty Principle.

\medskip
If, as is standard, the information $I$ is related to the
number of possible states $N$ by the equation $I =
\log_2N$, then the Bekenstein Bound on the amount of
information coded within a sphere of radius R
containing total energy E is 

$$I \le 2\pi ER/(\hbar c\ln
2)\eqno(2.1)$$

\noindent
or, expressing the energy in mass units
of kilograms, 

$$I\le 2.57686 \times 10^{43}\left(
{M}\over {\rm 1\, kilogram}\right)\left( {R}\over
{\rm\, 1\, meter}\right)\rm\, bits\eqno(2.2)$$

\medskip 
For example, a typical human being has a
mass of less than 100 kilograms, and is less than 2
meters tall.  (Thus such a human can be inscribed in a
sphere of radius 1 meter.)   Hence, we can let M equal
100 kg  and R equal 1 meter in formula (2.2) obtaining

$$I_{human}\le 2.57686 \times 10^{45} \rm\,
bits\eqno(2.3)$$

\noindent
as an upper bound to the number of bits
$I_{human}$ that can be coded by any physical entity
the size and mass of a human being. 

\medskip
Let me give an elementary {\it plausibility
argument} for the Bekenstein Bound $(2.1)$.  This
argument will be nonrigorous.  (A completely rigorous
proof would involve too much quantum field theory to be
feasible in this book.)  The Uncertainty Principle tells
us that 

$$\Delta P\Delta R \geq \hbar\eqno(2.4)$$

\noindent
Where $\Delta P$ is the ultimate limit in knowledge of
the momentum and $\Delta R$ is the limit in
knowledge of the position.  (Alternatively, the
inequality (2.4) expresses the minimum size of a phase
space division.)  Thus, if the total momentum is less
than $P$ and the system is known to be inside a region
of size $R$, then the phase space of the system must
be divided into no more than $PR/\Delta P\Delta R =
2\pi PR/h$  distinguishable subintervals.  This means
that the number of distinguishable states $n$ is
bounded above by $2\pi PR/h$.  Since for any particle,
$P \leq E/c$, where $E$ is the total energy of the
system including the system's rest mass, with
equality holding only if the system is moving at the
speed of light, we have

$$I = \log_2 n \leq {n\over \ln2} \leq 2\pi\left({E\over
c}\right)\left({R\over h\ln2}\right) \leq {2\pi
ER\over\hbar c\ln 2}$$ which is the Bekenstein Bound
(2.1).  (Additional complications like particle
substructure, and the fact that the system is in three
dimensions rather than one are implicitly taken into
account by the fact that $\log_2 n$ is very much less
than n, for large n.  As I said, the above derivation is
nonrigorous.)

\medskip 
An upper bound to the information processing rate can
be obtained [1] directly from the Bekenstein Bound by
noting that the time for a state transition cannot be
less than the time it takes for light to cross the sphere
of radius R, which is 2R/c.  Thus 

$$\dot I \le
{I\over{2R/c}} \le {\pi E\over\hbar\ln 2} = 3.86262
\times 10^{51}\left( {M}\over {\rm1\, kilogram}\right)
\rm\, bits/sec\eqno(2.5)$$ where the dot denotes the
proper time derivative.  By inserting 100 kilograms for
the value of M in inequality (2.5), we obtain an upper
bound for the rate of change of state of a human being,
$\dot I_{human}$:  

$$\dot I_{human} \le 3.86262 \times
10^{53}\rm\, states/sec\eqno(2.6)$$ 

\medskip The
significant digits in the RHS of inequalities (2.2), (2.3),
(2.5), and (2.6) have to be taken with a grain of salt. 
The digits correctly express our knowledge of the
constants $c$ and $\hbar$.  But the Bekenstein Bound is
probably not the least upper bound to either $I$ or
$\dot I$; Schiffer and Bekenstein have recently
shown [2] that the Bekenstein Bound probably
overestimates both  $I$ and $\dot I$ by a factor of at
least 100. 

\medskip
Strictly speaking, (2.5) only applies to a
single communication channel [3], but it probably [4]
applies even to multichannel systems if the need to
merge the information from various channels is taken
into account.  However, if the latter is not taken into
account, the number of channels is certainly limited
by the number of states given by (2.1), and so an
extremely conservative upper bound is ${dI\over d\tau} 
\le e^{I_{max}^B }\dot I_{max}^B$ (J. D. Bekenstein,
private communication).

\medskip
A human being --- indeed, any object existing in the
current universe --- actually codes far less than
quantum field theory would permit it to code.  For
example, a single hydrogen atom, if it were to code as
much information as permitted by the Bekenstein
Bound, would code about $4\times10^{6}$ bits of
information, since the hydrogen atom is about one \AA
ngstr\o m in radius, and has a mass of about
$1.67\times 10^{-27}$ kilograms.  So a hydrogen atom
could code about a megabyte of information, whereas it
typically codes far less than a bit.  The mass of the
hydrogen is not being used efficiently!

\medskip
If we take the radius to be that of a proton ($R =
10^{-13}$ cm.), then the amount of information that
can be coded in the proton is only 44 bits!  This is
remarkably small considering the complexity of the
proton --- three valence quarks, innumerable sea
quarks and gluons --- so complex in fact that we have
been unable to compute its ground state from first
principles using the Standard Model even when we use
our most advanced supercomptuers.  Bekenstein has
used this extremely small number of possible states in
the proton to constrain the number of possible quark
fields that could be present in the quark sea.

\medskip
In the early universe, where there are particle
horizons, and also for black holes, the Bekenstein
Bound in the form

$$I = {S\over \ln2} \leq {A\over 4L_P^2\ln2} = {\pi R^2
\over L_P^2 \ln2}\eqno(2.7)$$

\noindent
is appropriate, where $S$ is the total entropy in a
causally connected region inside a 2-sphere of radius
$R$ and surface area $A$, where $L_P$ is the Planck
length.  The Bekenstein Bound in the form (2.7) can be
easily derived from (2.1) as follows.

\medskip
If $R=2GE/c^4$, then a black hole forms, enclosing the
information, and {\it in asymptotally flat space} we
cannot get any more energy into a sphere of radius $R$
than this. Thus

$$I\leq { 2\pi ER\over \hbar c\ln2} =
2\pi\left({Rc^4\over 2G}\right){R\over \hbar c\ln2} =
4\pi R^2\left({c^3/G\hbar}\over 4\ln2\right)$$

But $c^3/G\hbar = L_P^{-2}$ and $A = 4\pi R^2$, so
we get (2.7).  However, the formation of a black hole
implies that there are event horizons, which means by
definition that the final singularity cannot be an
omega point.  That is, inequality (2.7) applies if and
only if the information corresponding to life is
restricted to a part rather than the entire universe.  

\medskip
Bekenstein has noted [5] that when a region in the
early universe with its particle horizons has a radius of
the order of a Planck length $L_P$, the entropy and
information must be of order one or less, from which
he concludes that the initial singularity does not
exist.  I would instead interpret this result (which I
believe to be correct) as implying that the initial
Friedmann singularity is {\it unique}; there is no
information in the initial singularity.  So $I = S = 0$ at
the initial singularity, and thus there is no
contradiction with the RHS of (2.7) going to zero as
$R\rightarrow 0$.  I shall show what this implies in
section 3.

\medskip
  Ellis and Coule [6] argue that, in {\it any} closed
universe near the {\it final} singularity, (2.7) is
still the correct form of Bekenstein's Bound, with $R$
being the scale factor of the universe, and thus
$R\rightarrow 0$ means $I\rightarrow 0$, which
obviously rules out  $I\rightarrow +\infty$ as
$R\rightarrow 0$, that holds if the Omega
Point Theory is true.  I shall show in Section H that if
(2.1) rather than (2.7) is used, we can have $I
\rightarrow +\infty$ as $R\rightarrow 0$, provided
event horizons disappear.

\medskip
But Ellis and Coule are wrong; (2.7) cannot be the
correct form near the final singularity in a closed
universe without event horizons because, if it were,
then it would imply a global and universal violation of
the Second Law of Thermodynamics when the radiation
temperature reaches a mere $5\times 10^4$ GeV,  far
below the Planck energy of $10^{19}$ GeV, and even far
below the unification temperature where we think
the Bekenstein Bound and the Second Law both apply.

\medskip
To see this, write $S = {\cal S}R^3$, where ${\cal S}$
is the entropy density, and let $R_0$ and $T_0$ be the
scale factor and radiation temperature today.  Using
$R = R_0T_0/T$, (2.7) implies the following upper
bound to the future universal temperature $T$:

$$T\leq {\sqrt\pi T_0\over \sqrt{{\cal
S}_0R_0L_p^2}}\eqno(2.8)$$

We have ${\cal S}_0 = 2.9\times 10^3$ cm$^{-3}$ from
equation $(B.17)$ of Section B of [9].  (See also [7, p.
273].  Note that applying $(B.17)$ to $(2.8)$ requires
leaving out the factor $\ln2$.)  Also, $T_0 =
2.726\rm\,^\circ K = 2.349 \times 10^{-13}\, GeV$. 
These numbers give

$$T \leq 5.3 \times 10^4 \rm\, GeV\eqno(2.9)$$

\noindent
if $R_0 = 3\rm\, gigaparsecs$ (the Hubble
distance) and $T \leq 3\times 10^3$ GeV if $R_0 =
1\rm\, teraparsec$, the lower bound in Section B of
[9].  If $R_0 = 10\rm\, teraparsecs$, the upper bound in
Section B of [9], then $T \leq 1\times 10^3$ GeV, which
is the energy the LHC is designed to probe.  Surely
quantum mechanics and the Second Law are valid at
this energy, even in the collapsing phase of a closed
universe.  I shall show that this is in fact the case in
Section 6.

\bigskip
{\bf b. Universe NOT Planck-sized at Planck Time}
\bigskip
To show that the universe must have been much
larger than the Planck Lenght at the Planck time, let us
suppose the early universe was radiation dominated and
topologically
$S^3$.  Since we know that it would have had to be
isotropic and homogeneous, it would be FRW and the
scale factor
$R(t)$ would evolve as

$$R(t) = R_{max}\left({{2t\over R_{max}} - {t^2\over
R_{max}^2}}\right)^{1/2}$$

\noindent
where $t$ is proper time, and $R_{max}$ is the scale
factor at maximum expansion.  Putting in $t=
L_{Pk}/c$ where $L_{Pk}$ is the Planck length, and
requiring $R( L_{Pk}/c)= L_{Pk}$ gives

$$R_{max} =  L_{Pk}$$

That is, the universe's maximum size is the Planck
length, in gross contradiction to observation.

\medskip

If we assume that the CBR radiation has been present
since the Planck time, which is to say that the
universe's expansion has been adiabatic since the
Planck time, then since in all cosmological models the
radiation density $\rho \propto 1/R^4$, we have

$$ R(L_{Pk}/c) = ({\rm Hubble\,\,
Distance})\left({\rho_{today}\over\rho_{Pk}}\right)
^{1/4} \approx10^{-4}\,{\rm cm}$$

\noindent
if we make the most natural assumption that the CBR
had the Planck density $\rho_{Pk}$ at the Planck time.
I shall show how the universe attains its ``unnatural''
enormous size in Section 6.

\bigskip
{\bf c. FRW Universe does NOT admit a U(1) gauge field,
like electromagnetism}

\bigskip
It has been know by relativists for many years that
a non-zero electromagnetic field --- a U(1) YM field ---
cannot exist in a FRW universe.  A simple proof is as
follows.  Equation (5.23) of Misner, Thorne and
Wheeler ([8], p. 141) gives the stress energy tensor for
the EM field in an orthnormal basis, in particular
$T^{{\hat0}{\hat j}} = (\vec{E} \times
\vec{B})^{\hat j}/4\pi$, which equals zero since in FRW
there can be no momentum flow.  Thus $\vec{B}$ must
be a multiple of $\vec{E}$, so set $\vec{B} = a\vec{E} =
aE\hat{x}$.  Computing the diagonal components of
$T^{\mu\nu}$ gives $T^{{\hat0}{\hat0}} =
E^2(1+a^2)/8\pi \equiv \rho$, and $T^{\hat{x}\hat{x}} =
-\rho = -T^{\hat{y}\hat{y}} = -T^{\hat{z}\hat{z}}$.  But
for FRW isotropy requires $T^{\hat{x}\hat{x}} =
T^{\hat{y}\hat{y}} =T^{\hat{z}\hat{z}}$, so $\rho =
({\vec E}^2 + {\vec B}^2/8\pi = 0$, which implies ${\vec
E} ={\vec B} = 0$).  However, any non-Abelian YM field
with an SU(2) normal subgroup {\it can} be non-zero in
a closed FRW, basically because SU(2) {\it is} a
homogeneous and isotropic 3-sphere.

\medskip
The fact that the FRW universe cannot admit an
electromagnetic field is ignored in standard
cosmology texts.   What is done is to assume that the
CMBR obeys the simple equation of state $p =
{1\over3} \rho$, and that the stress energy tensor is
simply the perfect fluid stress energy tensor.  The fact
that the CMBR, if it indeed {\it is} electromagentic
radiation, is ignored, or more precisely, is assumed to
result from some complicated averaging scheme which
is never spelled out in detail.  I shall argue throughout
this paper that the most natural interpretation of this
fact is that the CMBR is {\it not} a U(1) gauge field ---
it is {\it not} an electromagnetic field --- and this
this extraordinary claim has experimental
consequences, and amazingly, this extraordinary claim
is consistent will all observations to day of the
CMBR.  I shall also argue that just as the CMBR was
first seen in CN absorption lines, so the non-EM
nature of the CMBR has actually already been seen in
ultra high energy cosmic rays.

\bigskip
\centerline{\bf References}
\bigskip

\item{[1]} Bekenstein, J. D. 1981. {\it Physical
Review Letters} {\bf46}: 623. 
 
\medskip
\item{[2]} Schiffer, M. and J. D.
Bekenstein 1989.  {\it Physical Review} {\bf D39}:
1109; and {\it Physical Review} {\bf D42}: 3598.  

\medskip
\item{[3]} Bekenstein, J. D. 1988.  {\it Physical Review}
{\bf A37}: 3437.

\medskip
\item{[4]} Bekenstein, J. D. 1984. {\it Physical Review}
{\bf D30}: 1669.

\medskip
\item{[5]} Bekenstein, J. D. 1989.  {\it International 
Journal of Theoretical Physics} {\bf 28}: 967.

\medskip
\item{[6]} Ellis, G. F. R. and D.H. Coule 1992. ``Life at the
End of the Universe,'' University of Cape Town preprint.

\medskip
\item{[7]} B\"orner, G. 1992.  {\it The Early Universe}
Berlin: Springer.

\medskip
\item{[8]} C.W. Misner, K.S. Thorne, and J.A Wheeler
1973. {\it Gravitation} (San Francisco: Freeman)
 
\medskip
\item{[9]} Frank J. Tipler 1994. {\it The Physics of
Immortality} (New York: Doubleday).

\vfill\eject

\font\bigtenrm=cmr10 scaled\magstep5

\centerline{\bf{\bigtenrm 3. The Spectra of Gauge
Fields }}
 \bigskip
\centerline{\bf{\bigtenrm in a FRW Background}}

\bigskip
\bigskip
It has long been known (e.g, [6], p. 72; [7], p. 515) that
the Rayleigh-Jeans long wavelength limit of the Planck
distribution can be produced non-thermally, since this
limit is simply an expression of the number of wave
modes allowed in a given volume.  However, I shall
show that the spectral distribution of a quantized
gauge Bose radiation field in an exact FRW cosmology
must also necessarily follow a Planck distribution. 
More generally, any radiation field (defined as matter
whose energy density is inversely proportional to the
fourth power of the scale factor in a FRW cosmology)
will necessarily obey the Wien displacement law,
irrespective of whether it is quantized or what
statistics the field obeys.

\medskip
My derivation of the Planck distribution without
assuming thermal equilibrium is analogous to
Hawking's deriviation of a Planckian distribution for
the emission of radiation from a black hole.  In
Hawking's original calculation, no assumption of
thermal equilibrium was made initially, but he
discovered that the black hole radiation emission was
Planckian, with the black hole surface gravity playing
the role of the temperature.  I shall show that in a FRW
cosmology, a quantized gauge boson field must also
have a Planck spectrum, with the quantity $\hbar c/R$,
where $R$ is the radius of the universe, playing the
role of temperature.  However, because of the isotropy
and homogeneity of the FRW cosmology, there is no
possibility of interpreting this quantity as a
temperature.

\bigskip
\bigskip
{\bf a. PROOF THAT {\it ALL} CLASSICAL
GAUGE FIELDS NECESSARILY} 
\settabs 9 \columns
\+&{\bf OBEY A WIEN DISPLACEMENT LAW IN A FRW
UNIVERSE}\cr
\bigskip
\bigskip

I shall first show that the spectral distribution of
radiation --- that is, any field whose energy density
is inversely proportional to the fourth power of the
radius of the universe --- in any Friedmann-Robertson
-Walker (FRW) cosmology necessarily obeys the Wien
displacement law in the form

$$ I(\lambda, R) = {f(\lambda/R)\over R^5} =
{\phi(\lambda/R)\over \lambda^5}\eqno(3.1)$$

\noindent
where $R = R(t)$ is the FRW scale factor at any given
time $t$, $\lambda$ is the wavelength, $I(\lambda, R)$
is the monochromatic emissive power of the radiation,
and $f(\lambda/R)$ and $\phi(\lambda/R)$ are unknown
functions of the single variable $\lambda/R$.  Notice
that in the form $(3.1)$, the reciprocal of the scale
factor has replaced the temperature $T$.  The
temperature does not appear in the form $(3.1)$, and
no assumption of thermodyamic equilibrium will be
used in the derivation of $(3.1)$.  That is, the
spectral distirbution $(3.1)$ will apply no matter what
the thermodynamic state of the radiation is; it will
even be consistent with the radiation being at
absolute zero.

\medskip
Recall that in the standard derivation of the Wien
displacement law, the first step is to establish
Kirchhoff's law, which yields the fact that the
intensity $I$ of the radiation at a given wavelength
$\lambda$ depends only on $\lambda$ and the absolute
temperature.  In the standard derivation of Kirchhoff's
law, the assumption of thermal equilibrium is required
to establish this.  If we have a single radiation field in
a FRW cosmology, then $I = I(\lambda, R)$ --- the
intensity at a given wavelength depends only on the
wavelength and the scale factor --- because there are
no other variables in the cosmology upon which the
intensity of radiation could depend.

\medskip
Second, we recall that in a closed universe, the number
of modes $\cal N$ is constant under the expansion:

$${\cal N} = {R\over \lambda} = {R'\over \lambda'}
\eqno(3.2a)$$

\noindent
where the primes denote the quantities at some
other time.  Equation $3.2a)$ can be re-written

$${R'\over R} = {\lambda'\over \lambda} \eqno(3.2b)$$

An alternative calculation following [1], pp.
777--778 shows that in addition the same relation
between the wavelengths and expansion factors also
hold infinitesimally: $ {d\lambda/ R)} =
{d\lambda'/ R')}$, or

$${d\lambda'\over d\lambda} = {R'\over R}\eqno(3.2c)$$

During the expansion, the energy density $U$ of a
radiation dominated universe also changes.  We have

$$dU = \left({4\over c}\right) I(\lambda, R) d\lambda
\eqno(3.3)$$

The energy density of any gauge field satisfies

$${dU\over dU'} = {\left({R'\over
R}\right)}^4\eqno(3.4)$$

Thus\ combining $(3.3)$ and $(3.4)$ gives

$${dU\over dU'} = {{\left({4\over c}\right) I(\lambda, R)
d\lambda}\over \left({4\over c}\right) I(\lambda', R')
d\lambda'} = {\left({R'\over R}\right)}^4 \eqno(3.5)$$

\noindent
which gives upon solving for $I(\lambda, R)$ while
using $(3.2b)$ and $(3.2c)$:

$$I(\lambda, R) = {\left({R'\over R}\right)}^4
{d\lambda'\over d\lambda}I(\lambda', R') = {\left({R'\over
R}\right)}^5I({\lambda R'\over R}, R') =
{\left({\lambda'\over \lambda}\right)}^5I({\lambda
R'\over R}, R')\eqno(3.6)$$

As in the usual derivation of the Wien displacement
law, we note that since the LHS of equation $(3.6)$
does not contain the variables $R'$ or $\lambda'$,
neither can the RHS.  Thus $(3.6)$ can be written

$$ I(\lambda, R) = {f(\lambda/R)\over R^5} =
{\phi(\lambda/R)\over \lambda^5} \eqno(3.1)$$

which is the Wien displacement law.  Notice that if
there were several non-interacting radiation fields
present, then each would satisfy the Wien
dispalcement law, but possibily with different
functions of $R$, since we are not assuming thermal
equilibrium.

\medskip
The maximum of the distribution $(3.1)$ is obtained by
differentiating the first form of $(3.1)$ with respect
to $\lambda$:

$${dI(\lambda, R)\over d\lambda}\bigg\vert_{\lambda
= \lambda_m} = {d\over d\lambda}{f(\lambda/R)\over
R^5}\bigg\vert_{\lambda
= \lambda_m} = {1\over R^6}f'(\lambda_m/R) = 0$$

\noindent
which tells us that the wavelength $\lambda_m$ of the
maximum of the distribution satisfies 

$${\lambda_m\over R} = {\rm constant} \eqno(3.7)$$

\medskip
Of course, the above calculation is meaningful only
if there {\it is} a maximum. The Rayleigh-Jeans law
obeys the Wien dispalcement law, and has no
maximum.  But the Rayleigh-Jeans law also suffers
from the ultraviolet divergence, and so is unphysical. 
The actual distribtion $I(\lambda, R)$ must either have
a maximum, or asymptotically approach a limiting
value as $\lambda \rightarrow 0$.

\bigskip
\bigskip
{\bf b. PROOF THAT {\it ALL} QUANTIZED GAUGE FIELDS
NECESSARILY}  
\+&{\bf HAVE A PLANCKIAN SPECTRUM IN A FRW
UNIVERSE}\cr
\bigskip
\bigskip

I shall now show that if the radiation field of Section
1 is in addition a quantized gauge boson gas, the
spectral distribution will follow the Planck
distribution irrespective of whether the gas is in
thermal equilibrium.  The key idea will be to follow
Planck's original derivation of his Law [2, 3], which
remarkably did NOT assume that the gas was at a
maximum of the entropy (i.e., did not assume
equilibrium), though, as is well-known, he did assume
in effect assume the energies of the gas particles were
quantized, and that these particles obeyed Bose
statistics.  As in the derivation of the Wien
displacement law, the reciprocal of the scale factor of
the FRW cosmology will replace the temperature in the
Planck distribution law.

\medskip
The first part of the derivation will be the same as
the standard derivation of the Planck distribution.

\medskip
Let us define the following quantities:

\medskip
$g_s\, \equiv$ number of modes in the $s$ energy
level;

\medskip
$n_s\, \equiv$ number of particles in the $s$ energy
level;

\medskip
$\epsilon\, \equiv$ the energy of the $s$ energy level;
and 

\medskip
$Q \, \equiv n_s/g_s\, =$ the occupation index.

\medskip
For a boson gas, the number of distinct arrangements is

$$P_s = {{(n_s + g_s
-1)!}\over{n_s!(g_s-1)!}}\eqno(3.8)$$

The number of distinct arrangements $P$ for all energy
levels is thus

$$P =
\prod_{s=1}^{s_{max}}P_s
= \prod_{s=1}^{s_{max}}{{(n_s + g_s -1)!}\over {n_s!(g_s
-1)!}}\eqno(3.9)$$

The information in this collection of
possible arrangements is

$$I \equiv \log P = \sum_{s=1}^{s_{max}}\lbrack
\log(n_s + g_s - 1)! - \log n_s! - \log(g_s -1)!\rbrack
\eqno(3.10)$$

If we assume $g_s \gg 1$, and  use Stirling's formula,
$(3.10)$ becomes

$$I = \sum_{s=1}^{s_{max}} \lbrack(n_s + g_s)\log(n_s
+ g_s) - n_s\log n_s - g_s\log g_s\rbrack = $$

$$I = \sum_{s=1}^{s_{max}} g_s\left[\left(1+ {n_s\over
g_s}\right)\log\left(1 + {n_s\over g_s}\right) -
{n_s\over g_s}\log{n_s\over g_s}\right]\eqno(3.11)$$

\noindent
where each term in the sum will be denoted $I_s$, the
information in each energy level.

Now we know by the Bekenstein Bound that 

$$I \leq {\rm constant}(RE)$$

In the situation of perfect isotropy and homogeneity
this must apply for each state $s$ independently, and
for each mode.  Since the information per mode can
depend only on the scale factor $R$ --- in the FRW
universe there is no other possible function for the
information per mode to depend on --- and since the
Bekenstein Bound gives a linear dependence on $R$ for
all values of the scale factor, the inequality can be
replaced by an equality:

$$ d(I_s/g_s) = {\cal T}R\epsilon_sd(n_s/g_s )\
=
{{\partial(I_s/g_s)}\over{\partial(n_s/g_s)}}
d(n_s/g_s)\eqno(3.12)$$

\noindent
where $\cal T$ is a constant to be determined. 
Equation $(3.12)$ can be written

$${{\partial(I_s/g_s)}\over{\partial(n_s/g_s)}} =
{\cal T}\epsilon_sR\eqno(3.13)$$

From equation $(3.11)$ we have 

$${I_s\over g_s} = \left(1+ {n_s\over
g_s}\right)\log\left(1 + {n_s\over g_s}\right) -
{n_s\over g_s}\log{n_s\over g_s}$$

\noindent
and so substituting for simplicity $Q \equiv n_s/g_s$ 
we can evaluate the partial derivative in $(3.13)$:

$${{\partial(I_s/g_s)}\over{\partial(n_s/g_s)}} =
{d\over dQ}\left[(1+Q)\log(1 +Q) - Q\log Q\right] =
\log\left({{1+Q}\over Q}\right) = \epsilon_s{\cal T}R$$

\noindent
which can be solved for 

$$n_s = {g_s\over{\exp(\epsilon_s{\cal T}R) -1}}
\eqno(3.14)$$

As is well-known, the infinitesimal number of modes
$dN$ in a volume $V$ in the frequency interval
$d\omega$ is

$$dN = {{V\omega^2d\omega}\over{\pi^2c^3}}$$

\noindent
so the energy per mode is

$$dE = \hbar\omega n_s/g_s = \hbar\omega dN/g_s = 
{{\hbar\omega V\omega^2d\omega}\over{\pi^2c^3
(\exp(\hbar\omega{\cal T}R) -1)}}$$

\noindent
which yields a spectral energy density $dU = dE/V$ of

$$dU = {{\hbar\omega^3d\omega}\over{\pi^2c^3
(\exp(\hbar\omega{\cal T}R) -1)}}\eqno(3.15)$$

Using $I(\lambda,R) = (4/c)dU$ we obtain

$$I(\lambda,R) = {{2\pi c^2h}\over{\lambda^5
(\exp({\cal T}chR/\lambda) - 1)}}\eqno(3.16)$$

Equation $(3.15)$ can be integrated over all frequencies
from 0 to $+\infty$  to give at total energy density

$$U = {{\pi^2}\over{15\hbar^3c^3}}\left({1\over{{\cal
T}R}}\right)^4\eqno(3.17)$$

In a radiation dominated closed FRW universe, we have
(e.g. [1], p. 735)

$$U = {{3R^2_{max}c^4}\over{8\pi GR^4}}\eqno(3.18)$$

\noindent
where $R_{max}$ is the scale factor at maximum
expansion.

\medskip
Equating $(3.17)$ and $3.18)$  yields the constant $\cal
T$:

$${\cal T} = \left({8\pi^3\over{45}}\right)^{1/4}
\left(L_{Pk}\over
R_{max}\right)^{1/2}\left(1\over{\hbar
c}\right)\eqno(3.19)$$

If we integrate equation $(3.16)$ over all $\lambda$ to
obtain the total information in the universe, we obtain 

$$I_{Total} = {2\pi^4\over 15}\left({45\over 8\pi^3}
\right)^{3/4}\left[{R_{max}\over L_{Pk}}\right]^{3/2}
\approx 4\left[{R_{max}\over
L_{Pk}}\right]^{3/2}\eqno(3.20)$$

This is independent of the scale factor of the universe
$R$ --- so the information in the gauge field does not
change with time  (a Planck distribution is unchanged
by the universal expansion), but nevertheless it is
non-zero, which may be contrary to expectation; one
might expect that the information in the universe is
zero in the absence of thermalization.

\medskip
Before addressing the origin of this information (hint:
the origin is obvious from equation $(3.20)$), let me
first point out that the number $(3.20)$ is completely
consistent with the Bekenstein Bound, which is 

$$I \leq  \left({2\pi\over \hbar
c}\right)\left(ER\right)\eqno(3.21)$$

Let me replace the constant $2\pi/\hbar c$ with the
constant $\cal T$, which I have assumed will give
equality:

$$I = {\cal T}\left(ER\right)\eqno(3.22)$$

\noindent
where $\cal T$ can be written

$$ {\cal T} = {1\over(90\pi)^{1/4}}\left({L_{Pk}\over
R_{max}}\right)^{1/2}\left({2\pi\over \hbar c}\right)
\eqno(3.23)$$

So, provided 

$$R_{max} \geq L_{Pk}\eqno(3.24)$$

\noindent
we will have 

$${\cal T} < {2\pi\over \hbar c} \eqno(3.25)$$

\noindent
and thus the Bekenstein Bound will hold.

\medskip
If we happened to be in a universe in which $R_{max}
< L_{Pk}$, then the crucial approximation $g_s \gg 1$
and the use of Stirling's formula,which allowed me to
re-write $(3.10)$ as $(3.11)$, would no longer be
valid, and we would obtain a different $\cal T$,
consistent with the Bekenstein Bound in this universe.

\medskip
 Thus the origin of the information in the universe is
the particular value for $R_{max}$, which is just one
of the possible values; as we shall see in Section 6,
there are an infinity of possible values, and our
``selection'' of the particular scale factor at maximum
expansion in our particular universe of the
``multiverse'' (a term which will be precisely defined
in Section 6),  generates the information.

\medskip
In fact, it is clear from $(3.23)$ that had we simply
imposed the requirement that the information be of
order 1 at the Planck radius, say by setting ${\cal T} =
1$, then $R_{max} \sim L_{Pk}$.  Alternatively,  let us
try to eliminate all reference to $R_{max}$, by the
dimensionally allowed replacement $kT \rightarrow
\hbar c/R$.  Then, using the standard expresssion
above for the energy density $U$ of radiation with
termperature $T$, we get

$$U = {\pi^2(kT)^4\over 15 (\hbar c)^3} = {\pi^2\over 15
(\hbar c)^3}\left({\hbar c\over R}\right)^4 =
{\pi^2\hbar c\over 15 R^4} = {3c^4\over
8\pi GR^2_{max} \sin^4\tau} = {3R^2_{max}c^4\over
8\pi GR^4}$$

\noindent
or, 

$${3R^2_{max}\over 8\pi G} = {\pi^2 \hbar c\over 15}$$

\noindent
which yields

$$R_{max} = {2\pi\over\sqrt15}L_{Pk}
\approx (1.6)L_{Pk}\eqno(3.26)$$

Which is to say, the scale factor at maximum
expansion is of the order of the Planck length.

\medskip
Another way we could try to set the constant $\cal T$
is to simply require that the information in the gauge
field be of the order of one bit.  (Since the expansion
is adiabatic, the radius will cancel out of the
calculation.)   Recall that the entropy of radiation is
given by

$$S = {4\pi^2 kV(kT)^3\over45(c\hbar)^3}\eqno(3.27)$$

Setting $kT = 1/{\cal T}R$ and using the volume of a
closed universe $V = 2\pi^2R^3$, we get 

$${\cal T} = {2\pi\over\hbar c}\left({\pi\over
45}\right)^{1/3}\left({k\over S}\right)^{1/3}
\eqno(3.28)$$

Setting $S/k = 1$ gives

$${\cal T} = {2\pi\over\hbar c}\left({\pi\over
45}\right)^{1/3} \approx {2.6\over \hbar
c}\eqno(3.29)$$

\noindent
which, as we have already seen, gives $R_{max}
\approx L_{Pk}$.

\medskip
Another alternative we could try would be to set the
number of quanta in the universe to be equal to one. 
Recall that the number $n$ of quanta is

$$n = {2\zeta(3)\over \pi^2c^3\hbar^3}\left(kT\right)^3
V = {2\zeta(3)\over
\pi^2c^3\hbar^3}\left({1\over  {\cal
T}R}\right)(2\pi^2R^3) = {4\zeta(3)\over (c\hbar {\cal
T})^3}$$

Setting $n=1$ gives 

$${\cal T}c\hbar =
(4\zeta(3))^{1/3} \approx 1.7$$

\noindent
which once again gives $R_{max} \approx L_{Pk}$.

\medskip
As Bekenstein has often pointed out, when horizons
are present, the correct ``size'' $R$ that really should
be put into the Bekenstein Bound is the horizon radius;
in the case of the early universe, this is the particle
horizon radius.  Let be show now that with this value
for ``size'', indeed the choice $(3.19)$ gives less than
one bit of information inside the horizon when the
particle horizon is the Planck Length.

\medskip
The equation for the particle horizon radius is

$$ds^2 = -dt^2 + R^2(t)d\chi^2 \equiv -dt^2
+(dR_{Particle})^2 = 0$$

\noindent
which when integrated (setting $R(0) = 0$) yields

$$R_{Particle} = t\eqno(3.30)$$

It is well known that for a radiation dominated FRW
closed universe the scale factor can be expressed in
closed form in terms of the proper time $t$:

$$R(t) = R_{max}\left( {2t\over R_{max}} - {t^2\over
R_{max}^2}\right)^{1/2}\eqno(3.31)$$

\noindent
which can be solved for the proper time $t$:

$$t = R_{max}\left(1- \sqrt{1-\left({R\over R_{max}}
\right)^2}\right)\eqno(3.32)$$

\noindent
valid for $0 < t \leq R_{max}$.  For $R\ll R_{max}$,
this is approximately

$$R_{Particle} \approx
{R^2(t)\over2R_{max}}\eqno(3.33)$$

The information inside the particle horizon in the early
universe is thus

$$I = {\cal T}(UR^3_{Particle})(R_{Particle}) =
{\cal T}UR^4_{Particle} 
={\cal T}\left({3R^2_{max}c^4\over 8\pi
GR^4}\right)\left({R^8\over 2R^4_{max}}\right) =$$
$$ =\left({8\pi^3\over
45}\right)^{1/4}\left({L_{Pk}\over
R_{max}}\right)^{1/2}\left({3R^4\over 8\pi
R_{max}^2L^2_{Pk}}\right)\eqno(3.34)$$

Putting $(3.33)$ into $(3.34)$ gives 

$$I = \left({8\pi^3\over
45}\right)^{1/4}\left({L_{Pk}\over
R_{max}}\right)^{1/2}\left({12R^2_{Particle}\over
L^2_{Pk}}\right)\eqno(3.35)$$

which is much, much less than one for
$R_{Particle}\leq L_{Pk}$.

\vfill\eject
\centerline{\bf{References}}
\bigskip
\noindent\hangindent=20pt\hangafter=1
\item{[1]}  Misner C  W, Thorne K S and 
Wheeler J A 1973  {\it Gravitation} (San Francisco:
Freeman).

\medskip
\noindent\hangindent=20pt\hangafter=1
\item{[2]} Planck, Max 1959 {\it The Theory of Heat
Radiation} (New York: Dover Publications).

\medskip
\noindent\hangindent=20pt\hangafter=1
\item{[3]} Jammer, Max 1966 {\it The Conceptual
Development of Quantum Mechanics} (New York:
McGraw-Hill), pp. 20--22.

\medskip
\noindent\hangindent=20pt\hangafter=1
\item{[4]} Leighton, Robert B. {\it Principles of Modern
Physics} (New York: McGraw-Hill), pp. 62--65;
328--335.

\medskip
\noindent\hangindent=20pt\hangafter=1
\item{[5]} Tipler F J 1994 {\it The Physics of
Immortality} (New York: Doubleday).

\medskip
\noindent\hangindent=20pt\hangafter=1
\item{[6]} Weinberg, Steven 1977 {\it The First Three
Minutes} (London: Fontana/Collins).

\medskip
\noindent\hangindent=20pt\hangafter=1
\item{[7]} Weinberg, Steven 1972 {\it Gravitation and
Cosmology} (New York: Wiley).

\vfill\eject

\font\bigtenrm=cmr10 scaled\magstep5
\centerline{\bf{\bigtenrm 4. Particle Production
Solution}}
 \bigskip
\centerline{\bf{\bigtenrm to the EYM Equation in a
FRW Universe}}

\bigskip
\bigskip

\centerline{\bf Exact Solution of EYM Equations
with Constant SU(2) Curvature}

\bigskip

The Yang-Mills field (curvature) is

$$W^{\mu\nu}_a = \partial^\mu W^\nu_a - \partial^\nu
W^\mu_a + gf_{abc}W^\mu_bW^\nu_c$$

\noindent
where $f_{abc}$ are the structure constants of the Lie
group defining the Yang-Mills field, the Latin index
is the group index, and we summation over all repeated
Latin indicies.  In the absence of all other fields
except gravity, the YM fields satisfy the equations [10,
p. 13]

$$\nabla\wedge W = 0$$

\noindent
and

$$\nabla\wedge ^*W = 0$$

\noindent
where $\nabla$ is the gauge and spacetime covariant
derivative.  The first equation is the Bianchi identity
for the YM fields, while the second is the Yang-Mills
equation.  It is obvious that a self-dual ($W=^*W$) or
anti-self-dual ($W=-^*W$ will automatically satisfiy
both equations if it satisfies one.

\medskip
In more conventional notation, the Bianchi identity is

$$D_\mu W^a_{\nu\lambda} + D_\nu W^a_{\lambda\mu}
+ D_\lambda W^a_{\mu\nu} =0$$

\noindent
where

$$D_\lambda W^a_{\mu\nu} = W^a_{\mu\nu;\lambda} -
f^a_{bc}A^c_\lambda W^b_{\mu\nu}$$

\noindent
with the semicolon denoting the spacetime
covariant derivative and $A^c_\lambda$ being the gauge
potential, in terms of which the gauge field
$W^a_{\mu\nu}$ can be expressed as

$$ W^a_{\mu\nu} = A^a_{\nu;\mu} - A^a_{\mu;\nu} +
f^a_{bc}A^b_\mu A^c_\nu$$

$$= A^a_{\nu,\mu} - A^a_{\mu,\nu} + f^a_{bc}A^b_\mu
A^c_\nu$$

\noindent
where the last equality is valid in any coordinate
basis; that is, if the spacetime covariant derivative is
expressed in a coordinate basis, the spacetime
covariant derivatives can be replaced by partial
deiviatives.  The same is true in the Binachi identify
for the gauge fields $W$.

\medskip
The Yang-Mills equation in more conventional notation
is [11, p. 12]:
$$D^\nu W^a\,_{\mu\nu} = 0 =
W^a\,_\mu\,^\nu\,_{;\nu} - f^a_{bc}A^c\,_\nu
W^b\,_\mu\,^\nu$$

 \medskip
The Lagrangian for the YM field is $L =
-(1/16\pi)W^{\mu\nu}_aW^a_{\mu\nu}$, and the
standard expression for the stress energy tensor
$T_{\mu\nu} = -2\delta L/\delta g^{\mu\nu} +
g_{\mu\nu}L$ yields

$$T^{YM}_{\mu\nu} =
{1\over4\pi}\left[W^a_{\mu\beta}W_{a\nu}^\beta -
{1\over4}g_{\mu\nu}W^a_{\alpha\beta}
W^{\alpha\beta}_a\right]$$

For any $T^{YM}_{\mu\nu}$, we have $T^\mu_\mu =
0$, and so for an isotropic and homogeneous universe,
any YM field must have $T_{\hat i \hat i} \equiv p =
{1\over3}T_{\hat t \hat t}$, where $T_{\hat t \hat t} 
\equiv \rho$ in any local orthonormal frame.  In other
words, any YM field satisfies in a FRW universe a
perfect fluid equation of state with adiabatic
index $\gamma =4/3$.

\medskip
However, very few Yang-Mills fields are consistent
with isotropy and homogeneity.  It is well-known that
a non-zero electromagnetic field --- a U(1) YM field ---
cannot exist in a FRW universe.  (Proof: eq. (5.23)
of [1], p. 141, gives the stress energy tensor for the
EM field in an orthnormal basis, in particular
$T^{{\hat0}{\hat j}} = (\vec{E} \times
\vec{B})^{\hat j}/4\pi$, which equals zero since in FRW
there can be no momentum flow.  Thus $\vec{B}$ must
be a multiple of $\vec{E}$, so set $\vec{B} = a\vec{E} =
aE\hat{x}$.  Computing the diagonal components of
$T^{\mu\nu}$ gives $T^{{\hat0}{\hat0}} =
E^2(1+a^2)/8\pi \equiv \rho$, and $T^{\hat{x}\hat{x}} =
-\rho = -T^{\hat{y}\hat{y}} = -T^{\hat{z}\hat{z}}$.  But
for FRW isotropy requires $T^{\hat{x}\hat{x}} =
T^{\hat{y}\hat{y}} =T^{\hat{z}\hat{z}}$, so $\rho =
({\vec E}^2 + {\vec B}^2/8\pi = 0$, which implies ${\vec
E} ={\vec B} = 0$).  However, any non-Abelian YM field
with an SU(2) normal subgroup {\it can} be non-zero in
a closed FRW, basically because SU(2) {\it is} a
homogeneous and isotropic 3-sphere.

\medskip
If the YM connection is a left invariant 1-form, that is,
if the connection is a Maurer-Cartan form, then the
resulting YM curvature will be given spatially by the
structure constants of the Lie group.  The YM curvature
will be

$$W^{\mu\nu}_a = gf_{abc}W^\mu_bW^\nu_c$$

\noindent
where
$$W^\mu_a = R^{-1}(t)\delta^\mu_a$$

\noindent
with $a = 1,\, 2,\, 3$ being the spatial indices
and $R(t)$ being the FRW scale factor.  It is easily
checked that the above expression for
$T^{YM}_{\mu\nu}$ gives $T^{\hat{j}\hat{j}} =
(1/3)T^{\hat{0}\hat{0}} \propto R^{-4}$ and all other
components zero, provided that $f_{abc}
=\epsilon_{abc}$, the structure constants for SU(2).

\medskip
It is clear that the above expression for the gauge
field is consistent only if $R(t) =$ constant.  The true
time dependent SU(2) gauge field is

$$ W^{\mu\nu}_a = \left[\epsilon_{abc} \delta^\mu_b
\delta^\nu_c \pm
{1\over2}\epsilon^{\mu\nu\alpha\beta}\epsilon_{abc}
\delta^b_\alpha\delta^c_\beta\right]{A\over R^2(t)}$$

\noindent
where $A$ is a constant, fixed by the Higgs field as I
shall show below.  The plus sign gives a self-dual
gauge field ($W^{\mu\nu}_a = +^*W^{\mu\nu}_a$), and
the minus sign gives an anti-self-dual field 
($W^{\mu\nu}_a =-^*W^{\mu\nu}_a$).

\bigskip
\centerline{\bf References}
\bigskip

\noindent\hangindent=20pt\hangafter=1
\item{[1]} C. W. Misner, K.S. Thorne, and J.A. Wheeler,
1973 {\it Gravitation} (Freeman: San Francisco).

\vfill\eject

\font\bigtenrm=cmr10 scaled\magstep5
\centerline{\bf{\bigtenrm 5. Particle Production by 
Instanton}}
 \bigskip
\centerline{\bf{\bigtenrm Tunnelling in a
FRW Universe}}

\bigskip
\bigskip

Since the Bekenstein Bound requires a unique initial
state, and since the only allowed non-zero initial field
is the isotropic and homogeneous SU(2) field, the initial
baryon number is necessarily zero; the Bekenstein
Bound thus {\it requires} a mechanism to produce a net
baryon number.  Baryogenesis requires satisfying the
three Sakharov conditions: 

\medskip\noindent 
(1) violation of baryon number B and
lepton number L conservation;

\medskip\noindent
(2) violation of C and CP invariance; and
 
\medskip\noindent
(3) absence of thermal equilibrium

\medskip
The SM has a natural method of baryogensis via the
triangle anomaly, which generates both baryon and
lepton number (B + L is not conserved but B - L is
conserved), and since the self-dual gauge field
generates fermions rather than anti-fermions, it
violates C.  The anomaly function
$^*W^a_{\mu\nu}W_a^{\mu\nu}$ can be written ${\bf
E}^a\cdot{\bf B}_a$.  Since ${\bf E}^a$ is odd under
parity while ${\bf B}_a$ is even, the anomaly function
will violate CP.  At zero temperature, all processes
will be effectively non-equilibrium process.  So
baryogenesis via the triangle anomaly at zero
temperature is the natural method of SM baryogenesis.

\medskip
The Standard Model violates CP perturbatively via the
complex phase in CKM matrix.  In the early universe,
this perturbative mechanism fixes whether fermions or
anti-fermions will be created via the triangle anomaly;
that is, it fixes the SU(2) gravitational sphaleron to be
a self-dual rather than an anti-self-dual solution to the
EYM equations.  At high temperatures, the triangle
anomaly will on average generate almost as many
anti-fermions as fermions, because in thermal
equilibrium the SU(2) gauge fields will be
anti-self-dual as often as self-dual.  Thus, in
equilibrium, the CP violating CKM complex phase acts
to surpress SM baryogeneis; the excess of fermions
over anti-fermions is surpressed by the Jarlskog
determinant factor.  As is well-known, this
surpression can wash out at 100 GeV any fermion
excess generated at higher temperature by other
mechanisms.  

\medskip
In the usual high temperature SM baryogenesis
calculation ([1], [2], [3], the baryon to photon ratio in
dimensionaless units , known (e.g. [4]) from
nucleosynthesis to be $\eta_{-9} = 1.0
\pm 0.15$, is too small by a factor of about $10^{-8}$,
because the net creation rate is suppressed by the
smallness of the CP violation in the CKM matrix
described above, even when the problem of washing out
any net baryon number is ignored.  These problems are
avoided in my proposed mechanism, for two reasons:
first, the only role played by the CP violation is to
select a self-dual rather than an anti-self dual field
as the unique field (in the absence of CP violation
there would be no unique SU(2) field; the actual
magnitude of the CP violation is irrelevant.  Second,
the baryon number generation is always at zero
temperature, so there will never be any anti-fermions
generated, and never any washout of fermions created. 
In fact, my model may have the opposite problem from
the usual electroweak high temperature model: my
model may tend to produce too many baryons relative to
the number of SU(2) pseudo-photons..

\medskip
The net baryon number generated by the SU(2)
sphaleron is given by [2; 5, p. 454]:

$$N = {-1\over{32\pi^2}}\int \sqrt{-g} d^4x
\,[{1\over2}\epsilon
^{\alpha\beta\mu\nu}W^a\,_{\alpha\beta}
W^b\,_{\mu\nu}(tr \,t_at_b)]\eqno(5.1)$$

I have set the $SU(3)$ gauge field to zero.  Once
again, this is required by uniqueness.  There are
uncountably many $SU(2)$ subgroups in $SU(3)$, but
thery are all in the same congugacy class.  A simple
proof of this is as follows (this simple proof was
pointed out to me by J. Bryan).

\medskip
Suppose $G$ is a subgroup of $SU(3)$ and $G$ is
isomorphic to $SU(2)$. Then the action of $SU(3)$ on
$C^3$ (complex Euclidean 3-space) induces a three
dimensional representation of $G$.  Since any
representation is the direct sum of irreducible
representations, this representation must be (1) the
unique irreducible representation of dimension three
(spin one representation), or (2) a direct sum of the
unique representation of dimension two (spin one half
representation) plus the one dimensional (trivial,
spin zero) representation, or (3) a direct sum of three
one dimensional (trivial) representations. (I use the
fact that $SU(2)$ has a unique irreducible
representation in each dimension). It cannot be (3)
since $G$ is a subgroup and so acts non-trivially. It
cannot be (1) since this representation is isomorphic
to the adjoint representation of $SU(2)$ on its Lie
algebra and so the negative of the identity element
acts trivially and so $G$ would not be a subgroup. 
Therefore the representation must be a direct sum of
the standard two dimensional representation and the
trivial representation.  Choose a unitary basis of $C^3$
so that the last factor is the trivial representation and
first two factors are the standard representation and
this change of basis will conjugate $G$ into the
standard embedding of $SU(2)$ into $SU(3)$. QED.  (As
an aside, note that we have the double cover $SU(2)
\rightarrow SO(3) \subset SU(3)$. The induced
representation on $SU(2)$ in this case will in fact be
the irreducible three dimensional one, but in this case
the subgroup is $SO(3)$, not $SU(2)$.)

\medskip
However, even though all $SU(2)$ subgroups are in the
same congugacy class, they are not physically
equivalent.  Each different $SU(2)$ subgroup is
generated by a different Lie subalgebra corresponding
to a different linear superpostion of gluons.  Each
such linear superposition is physically different.  Thus
there are uncountably many physically distinct $SU(2)$
subgroups of $SU(3)$, each capable of generating a
homogeneous and isotropic metric (since isomorphic
to the electroweak $SU(2)$ used above).  This means
the only way to have a unique $SU(3)$ field over a FRW
spacetime is to set {\it all} the gluons fields to zero.

\medskip
I have exhibited the two essentially unique vacuum
solutions to the ETM equations in Section 4; since, as I
have argued above, the self-dual is the unique solution
required by the SM, we take the plus sign:

$$ W^{\mu\nu}_a = \left[\epsilon_{abc} \delta^\mu_b
\delta^\nu_c \pm
{1\over2}\epsilon^{\mu\nu\alpha\beta}\epsilon_{abc}
\delta^b_\alpha\delta^c_\beta\right]{A\over R^2(t)}
\eqno(5.2)$$

Putting $(5.2)$ into $(5.1)$ gives

$$N = {1\over{32\pi^2}}\int {6A^2\over R^4} \sqrt{-g}
d^4x = {3\pi\over 8}A^2\int^t_0{dt\over R}\eqno(5.3)$$

The last integral is of course conformal time; net
fermion production is proportional to conformal time,
and this suggests that the most appropriate time
variable for quantum gravity is conformal time, since
conformal time and only conformal time measures the
rate at which something new is occurring to break
perfect symmetry: fermions are appearing in the pure
SU(2) gauge field.  This fact of the nature of the natural
time variable will be used in Section 6 to quantize the
gravitational field in the early universe.

\medskip
The reader should be aware that I have not shown that
my mechanism will in fact produce the correct
observed baryon to photon ratio; I have only argued
that my mechanism is not in principle subject to the
well-known limitations of SM electroweak
baryogenesis.  I conjecture that my mechanism will
produce the correct ratio; investigation of my
conjecture will be be subject of a subsequent paper.

\bigskip
\centerline{\bf References}
\bigskip

\noindent\hangindent=20pt\hangafter=1
\item{[1]} V.A. Kuzmin, V.A. Rubakov, and M.E.
Shaposhnikov, ``On Anomalous Electroweak
Baryon-Number Non-Conservaton in the Early
Universe,'' Phys. Lett. {\bf B155} (1985), 36--42.

\medskip
\noindent\hangindent=20pt\hangafter=1
\item{[2]} V.A. Rubakov, M.E. Shaposhinikov,
``Electroweak Baryon Number Nonconservation in the
Early Universe and in High-Energy Collisions,'' Physics
Uspekhi {\bf 39} (1996), 461--502.

\medskip
\noindent\hangindent=20pt\hangafter=1
\item{[3]} A.G. Cohen, D.B. Kaplan, and A.E. Nelson,
``Progress in Electroweak Baryogensis,'' Ann. Rev. Nucl.
Part. Sci. {\bf 43} (1993), 27--70.

\medskip
\noindent\hangindent=20pt\hangafter=1
\item{[4]} Craig J. Hogan, ``Cosmic Discord,'' Nature
{\bf 408} (2000), 47--48.

\medskip
\noindent\hangindent=20pt\hangafter=1
\item{[5]} S. Weinberg 1996 {\it The Quantum Theory
of Fields, Volume II} (Cambridge Univ. Press,
Cambridge).

\vfill\eject

\font\bigtenrm=cmr10 scaled\magstep5
\centerline{\bf{\bigtenrm 6. The Unique Quantized
FRW Universe}}

\bigskip
\bigskip

{\bf a. Conformal Time is the Unique Physical Time} 

\bigskip
In this section, I shall justify ignoring quantum
gravity effects --- quantum gravity fluctuations ---
in the very early universe.  I shall do this by
constructing a quantized FRW universe in which the
only field is a gauge field (actually a perfect fluid for
which $p = \rho/3$) and show that imposing the
boundary condition that classical physics hold exactly
at ``late times" (any time after the first minute)
implies that classical physics is good all the way into
the initial singularity.

\medskip
In standard quantum gravity, the wave function of the
universe obeys the Wheeler-DeWitt equation 

$$\hat {\cal H}\Psi = 0\eqno(6.1)$$

\noindent
where $\hat{\cal H}$ is the super-Hamiltonian
operator.  This operator contains the equivalent of
time derivatives in the Schr\"odinger equation.  I say
``the equivalent'' because quantum gravity does not
contain time as an independent variable.  Rather, other
variables --- matter or the spatial metric --- are used
as time markers.  In other words, the variation of the
physical quantities {\it is} time.  Depending on the
variable chosen to measure time, the time interval
between the present and the initial or final
singularity can be finite or infinite --- but this is
already familiar from classical general relativity.  In
the very early universe, conformal time measures the
rate at which particles are being created by instanton
tunnelling, that is it measures the rate at which new
information is being created.  Therefore, the most
appropriate physical time variable is conformal time,
and thus we shall select an appropriate combination of
matter and spatial variables that will in effect result
in conformal time being used as the fundamental time
parameter in the Wheeler-DeWitt equation.  Conformal
time is also the most natural physical time to use for
another reason:  The matter in the early universe
consists entirely of an $SU(2)$ gauge field, and the
Yang-Mills equation is conformally invariant; a gauge
field's most natural time variable is conformal time.

\medskip
Since the Bekenstein Bound tells us that the
information content of the early universe is zero, this
means that the only physical variable we have to
take into account is the scale factor $R$ of the
universe, and the density and pressure of the gauge
field.  So we only have to quantize the FRW universe for 
a radiation field, or equivalently, a perfect fluid for
which $p = \rho/3$.

\medskip
If matter is in the form of a perfect fluid, the action
$S$ in the ADM formalism can be written

$$S = \int (\Re + p)\sqrt{-g}\, d^4x = \int L_{ADM}
\,dt \eqno(6.2)$$

\noindent
where $p$ is the fluid pressure and $\Re$ is the
Ricci scalar.  If the spacetime is assumed
to be a Friedmann universe containing isentropic perfect
fluids, Lapchinskii and Rubakov [2] have shown the
canonical variables can be chosen $(R,\phi,s)$, where
$R$ is the scale factor of the universe, and $\phi,s$
are particular parameterizations of the fluid variables
called {\it Schutz potentials} [3].  The momenta
conjugate to these canonical variables will be written
$(p_R,p_\phi, p_s)$.

\medskip
The ADM Lagrangian in these variables can be shown to
be

$$L_ADM = p_RR' + p_\phi \phi' + p_ss' - N(H_g + H_m)
\eqno(6.3)$$

\noindent
where the prime denotes the time derivative,

$$H_g = - {p^2_R\over24R} - 6R \eqno(6.4)$$

\noindent
 is the purely gravitational super-Hamiltonian, and

$$H_m = N^2R^3[(\rho + p)(u^0)^2 +pg^{00}] = p^\gamma
_\phi R^{3(1-\gamma)} e^s \eqno(6.5)$$

\noindent
is both the coordinate energy density measured by a
comoving observer and the super-Hamiltonian of the
matter.  The momentum conjugate to $R$, the scale
factor of the universe, is

$$p_R = -{12RR'\over N}\eqno(6.6)$$

The constraint equation for the Friedmann universe is 
obtained by substituting $(6.3)$ -- $(6.5)$ into $(6.2)$
and varying the lapse $N$.  The result is the
super-Hamiltonian constraint:

$$0 = {\cal H} = H_g + H_m = -{p^2_R\over 24R} -6R +
p^\gamma_\phi R^{3(1-\gamma)}e^s \eqno(6.7)$$

When the perfect fluid is radiation the last term is
$H_m = p_\phi^{4/3}e^s/R$, and so if we choose the
momentum conjugate to the {\it true} time $\tau$ to be

$$p_\tau = p^{4/3}_\phi e^s \eqno(6.8)$$

\noindent
then the super-Hamiltonian constraint becomes

$$0 = {\cal H} = -{p_R^2 \over 24R} -6R + {p_\tau\over
R} \eqno(6.9)$$

The ADM Hamiltonian is obtained from $H_{ADM} =
p_\tau$, or 

$$H_{ADM} ={ p_R^2\over 24} + 6R^2\eqno(6.10)$$

\noindent
which is just the Hamiltonian for a simple harmonic
oscillator.

\medskip
The lapse $N$ is fixed by solving Hamilton's equation
$$\tau' = 1 = {\partial (N[H_g + H_m])\over \partial
p_\tau} ={N\over R} \eqno(6.11)$$

\noindent
which says that $N = R$; that is, {\it true} time is
just conformal time, which is why I have called it
$\tau$.

\medskip
If we quantize by the replacement $p_\tau \rightarrow
\hat p_\tau = -i \partial /\partial \tau$, and $p_R
\rightarrow \hat p_R = -i\partial /\partial R$, together
with a reversal of the direction of time
$\tau\rightarrow -\tau$ in the super-Hamiltonian
constraint $(6.9)$, the Wheeler-DeWitt equation
$(6.1)$ will then become (if we ignore factor ordering
problems) Schr\"odinger's equation for the simple
harmonic oscillator with mass $m = 12$, spring
constant $k=12$ and angular frequency $\omega = 1$:

$$i {\partial\Psi \over\partial \tau} = -{1\over
24}{\partial^2\Psi\over\partial R^2} + 6R^2\Psi
\eqno(6.12)$$

\bigskip
{\bf b. Consistency between Copenhagen and Many-Worlds
Interpretations}
\settabs 9 \columns
\+& {\bf Requires a Delta Function Initial Boundary
Condition}\cr

\bigskip
We need to find what boundary conditions to impose on
equation $(6.12)$.  The boundary condition that I
propose is the unique boundary condition that will
allow the classical Einstein equations to hold
exactly in the present epoch: that is, I shall require
that on the largest possible scales in the present
epoch, classical mechanics holds exactly.  To see how
to impose such a boundary condition, let us consider
the general one-particle Schr\"odinger equation

$$i\hbar{\partial\psi\over \partial t} = -{\hbar^2\over
2m}\nabla^2\psi + V(\vec x)\psi \eqno(6.13)$$

If we substitute ( [4], p. 280; [6], p. 51--52; [7])

$$\psi = {\cal R}\exp(i{\varphi}/h)\eqno(6.14)$$

\noindent
into (6.13), where the functions ${\cal R} = {\cal
R}(\vec x,t)$ and $\varphi = \varphi(\vec x,t)$ are real,
we obtain

$${\partial {\cal R}\over\partial t} = -{1\over2m}\left[
{\cal R}\nabla^2\varphi + 2\vec\nabla
{\cal R}\cdot\vec\nabla \varphi\right] \eqno(6.15)$$

$${\partial \varphi\over\partial t} = - {(\vec\nabla
\varphi)^2 \over 2m} - V  +\left(\hbar^2\over
2m\right){\nabla^2{\cal R}\over {\cal R}}\eqno(6.16)$$

Equation (6.16) is just the classical Hamilton-Jacobi
equation for a single particle moving in the potential

$$U = V  -\left(\hbar^2\over
2m\right){\nabla^2{\cal R}\over {\cal R}}\eqno(6.17)$$

Equations (6.16) and (6.17) are fully equivalent to
Schr\"odinger's equation (6.13),  and this way of
expressing  Schr\"odinger's equation, called the
Bohm--Landau Picture ([6], [7]), is the most
convenient formulation of QM when one wishes to
compare QM with classical mechanics.  The normals to
surfaces of constant phase, given by $\varphi(\vec x,t)
= \rm\, constant$, define trajectories: those curves
with tangents

$$\vec\nabla \varphi = {\hbar\over 2im}\ln\left(\psi
\over \psi^\ast\right) = Re\left[\left(\hbar\over
i\right)\ln\psi\right]\eqno(6.18)$$

The density of the trajectories is conserved, since this
density is given by $\rho = \psi \psi^\ast = {\cal R}^2$,
satisfying 

$${\partial\rho\over\partial t} + \vec\nabla\cdot
\left(\rho{\vec\nabla \varphi\over m}\right)
=0\eqno(6.19)$$

\noindent
which is just (6.15) rewritten.

\medskip
The surfaces of constant phase guide an infinite
ensemble of particles, each with momentum $\vec p =
m\vec\nabla \varphi$:  {\it all} the trajectories
defined by (6.18) are real in quantum mechanics.  In all
quantum systems, Many Worlds are present, though if
we make a measurement, we will see only one particle.
But we must keep in mind that in actuality, there are
infinitely many particles --- infinitely many histories
--- physically present.  The same will be true in
quantum cosmology.

\medskip
But we will be aware of only one universe in quantum
cosmology, so the requirement that classical
mechanics hold exactly in the large in the present
epoch can only mean that this single universe of which
we are aware must obey exactly the classical
Hamilton-Jacobi equation: that is, we must require
that

$${\nabla^2{\cal R}\over {\cal R}} = 0\eqno(6.20)$$

By requiring (6.20) to be imposed on the wave
function of the universe in the present epoch, I have in
effect unified the Many-Worlds Interpretation of
quantum mechancs with Bohr's version of the
Copenhagen Interpretation.  In what is universally
regarded as Bohr's definitive article on the
Copenhagen interpretation --- his paper ``Discussion
with Einstein on Epistemological Problems in Atomic
Physics'' in the P. A. Schilpp's {\it Albert Einstein:
Philosopher-Scientist} [5] --- Bohr never once claims
that the wave function must be ``reduced''; i.e., undergo
non-unitary evolution, nor does he ever claim that
macroscopic systems such as human beings are not
subject to the unitary time evolution of atomic
systems.  Bohr instead asserts: ``$\ldots$ it is
decisive to recognize that, {\it however far the
phenomena transcend the scope of classical physical
explanation, the account of all evidence must be
expressed in classical terms.}'' ([5], p. 209, Bohr's
italics) ``$\ldots$ This recognition, however, in no
way points to any limitation of the scope of the
quantum-mechanical description $\ldots$'' ([5], p.
211).  

\medskip
But quantum mechanics has unlimited validity
only if it applies equally to human-size objects as
well as to atoms, and thus the requirement that
accounts of phenomena expressed at the human size
level and larger must be in classical mechanical terms
can only mean that the larger objects must obey
classical and quantum laws simultaneously.  And this
is possible only if the human-size objects and larger
obey Schr\"odinger's equation and the classical H-J
equation simultaneously, which requires that the
boundary condition (6.20) hold.

\medskip
But it is only required to hold at the present epoch
(more precisely, after the first minute), and only on
the largest possible scale, that of the universe as a
whole.  In other words, the Copenhagen Interpretation is
to be regarded as something like the Second Law of
Thermodynamics: it applies only to large scale
systems, and it holds exactly only on the largest
possible scale, the scale of the universe as a whole. 
The Many-Worlds Interpretation holds always, just as
statistical mechanics holds always.  But boundary
conditions msut be imposed on the MWI to yield exactly
the Copenhagen Interpretation in the large, just
boundary conditions must be imposed in statistical
mechanics to yield exactly the Second Law in the
Thermodynamic Limit.

However, as we shall see, imposing (6.20) today will
yield a classical evolution from the initial singularity
to the present day, thus justifying the use of classical
field equations in the very early universe arbitrarily
close to the initial singularity, as I have done in
previous sections.

\medskip
If ${\cal R}$ is bounded above --- as it would be if
$\psi$ were a function in a Hilbert space --- equation
(6.20) requires $\nabla^2{\cal R}=0$.  This in turn
implies (since ${\cal R}$ is bounded above) that ${\cal
R}=\rm\, constant$.  But the only allowed way in
quantum mechanics to obtain ${\cal R}=\rm\,
constant$ is to extend the Hilbert space to a Rigged
Hilbert space (Gel'fand triple) that includes delta
functions.  For example, when $V=0$, a delta function
in momentum space yields ${\cal R}=\rm\,constant$,
and the plane wave, which indeed satisfies the
classical H-J equation, and indeed the trajectories
which are everywhere normal to the constant phase
surfaces are the straight lines with tangents
proportional to the momentum vector.

\medskip
It is important to emphasize that ${\cal R}=\rm\,
constant$ is going to yield a non-normalizable wave
function, and that the only allowed non-normalizable
wave function are indeed delta functions.  For,
as B\"ohm has pointed out [8], the most fundamental
expression of the wave function, the Dirac kets, are
themselves delta functions, so delta functions are
physically real states that are actually physically
more fundamental than the Hilbert space states.  So we
should not be surprised to find that the initial state of
the universe is one of the most fundamental states.

\medskip
The wave function of the universe $\Psi(R,\tau)$ in the
mini-superspace described above is a function of two
variables, the scale factor of the universe $R$ and the
conformal time $\tau$.

\medskip
If the initial boundary condition

$$\Psi(0,0) = \delta(R)\eqno(6.21)$$

$${\left[\partial\Psi(R,\tau)\over
\partial R\right]_{R=0}} = 0\eqno(6.22)$$

\noindent
is imposed, then the resulting wave function will have
classical H-J trajectories for $\tau >0$.  (Boundary
condition (6.22) is imposed in addition to the delta
function condition (6.21) for the following reason.  The
wave function is assumed to have no support for $R <
0$.  However, we cannot get this by imposng the DeWitt
boundary condition $\Psi(0,\tau) = 0$, because it
contradicts $(6.21)$.  But $(6.22)$ is sufficient for
self-adjointness of the SHO Hamiltonian on the
half-line $R \in [0,+\infty)$; see [1] for a discussion.) 
The wave function satisfying boundary conditions
$(6.21)$ and $(6.22)$ is just the Green's function $G(R,
\tilde R,\tau)$ defined on the entire real line for the
simple harmonic oscillator, with $\tilde R$ set equal
to zero.  The wave function is thus

$$\Psi(R,\tau) = \biggl[{6\over \pi L_P\sin \tau}
\biggr]^{1/2}\exp \biggl[{i6 R^2\cot \tau\over
L_P^2}- {i\pi\over 4} \biggr]\eqno(6.23)$$

\noindent
where $L_{P}$ is the Planck length.  This wave
function is defined only for a finite conformal time: $0
\leq \tau \leq \pi$.  (The initial and final singularities
{\it are} in the domain of the wave function!)

\medskip
Notice that the magnitude of the wave function
$(6.23)$ is independent of the scale factor of the
universe $R$.  Since the scale factor plays the role of
``spatial position'' in the simple harmonic oscillator
equation $(6.12)$, we have $\nabla^2R = 0$, and hence
from the discussion on phase trajectories above,
we see that the phase trajectories for the wave
function $(6.23)$ are all the classical trajectories for
a simple harmonic oscillator.  That is, the phase
trajectories are all of the form

$$R(\tau) = R_{max}\sin \tau\eqno(6.24)$$

\noindent
which are also all the classical solutions to the
Einstein field equations for a radiation-dominated
Friedmann universe.

\medskip
We can also show that the phase trajectories are given
by $(6.24)$ by direct calculation.  Since in the natural
units $L_P=1$, the phase $\varphi$ is $\varphi =
6R^2\cot\tau - {\pi\over4}$, we have $\nabla\varphi =
\partial\varphi/\partial R =12R\cot\tau$.  The tangents
are defined by $p_R =  \nabla\varphi$, which implies 

$${1\over R}{dR\over d\tau} = \cot\tau\eqno(6.25)$$

\noindent
using $(6.6)$, $N=R$, and $\tau\rightarrow -\tau$. 
The solutions to $(6.25)$ are $(6.24)$.  

\medskip
With the boundary condition $(6.21)$, {\it all} radii at
maximum expansion, $R_{max}$, are present; all
classical paths are present in this wave function.  We
thus see that, with the boundary condition $(6.21)$,
both the phase trajectories and the wave function
begin with an initial singularity and end in a final
singularity.  In other words, with this wave function,
the universe behaves quantum mechanically just as it
does classically.  The singularities are just as real in
both cases.  Conversely, we can run the calculation in
reverse and conclude that in a SHO potential with 
${\cal R}=\rm\, constant$, we see that the universal
wave function must have been initially a delta
function.

\bigskip
{\bf c. Solution to Flatness Problem in Cosmology}
\bigskip

Since $\rho(R(\tau)) = \psi \psi^\ast = {\cal R}^2$
measures the density of universes with radius $R$,
for normalizable wave functions, it implies the Born
Interpretation: the probability that we will find
ourselves in a universe with size $R$ is given by ${\cal
R}^2$.   Similarly, if ${\cal R}=\rm\, constant$, we are
equally likely to find ourselves in a universe with any
given radius.  However, since $R>0$, if we ask for the
relative probability that we will find ourselves in
a universe with radius larger than any given radius
$R_{given}$ or instead find ourselves in a universe with
radius smaller than $R_{given}$, we see that the
relative probability is one that we will find ourselves
in a universe with radius larger than $R_{given}$, since
$\int^\infty_{R_{given}}{\cal R}^2\,d {\cal R} = +\infty$
while $\int^{R_{given}}_0{\cal R}^2\,d {\cal R}$ is
finite.  Thus with probability one we should expect to
find ourselves in a universe which if closed is
nevertheless arbitrarily close to being flat.  This
resolves the Flatness Problem in cosmology, and we
see that we live in a flat universe because (1) the
Copenhagen Interpretation applies in the large,
or equivalently, because (2) the quantum universe began
as a delta function at the initial singularity, or
equivalently, because (3) classical physics applies on
macroscopic scales.

\medskip
Notice a remarable fact:  although the above
calculation was done using the Wheeler-DeWitt
equation, the same result would have been obtained if I
had done it in classical GR (in its Hamilton-Jacobi
form), or even done it in Newtonian gravity (in its
Hamilton-Jacobi form).  Just as one can do FRW
cosmology in Newtonian gravity, so one can also do
FRW cosmology in quantum gravity.  The conclusion is
the same in all theories: the universe must be flat. 
This conclusion does not, in other words, depend on the
value of the speed of light, or on the value of Planck's
constant.  In short, the flatness conclusion is robust!

\bigskip
{\bf d. Solution to the Standard Cosmological
Problems:}

\+& {\bf Homogeneity, Isotropy, and Horizon}\cr 

\bigskip
One might think that quantum fluctuations would
smear out the classical nature of spacetime near the
initial singualrity.  Let me now prove that this is
false; that in fact the consistency of quantum
mechanics with general relativity requires these
fluctuations to be suppressed.  It is not the quantum
fluctuations at the instant of their formation that
gives rise to an inconsistency, but rather how such
fluctuations would evolve in the far future. 
Fluctuations will in general yield mini--black holes,
and it is the evolution of black holes, once formed,
that give rise to inconsistencies.

\medskip
Astrophysical black holes almost certainly exist, but
Hawking has shown ([9]; [15], Section 7.3) that if black
holes are allowed to exist for unlimited proper time,
then they will completely evaporate, and unitarity will
be violated.  Thus unitarity requires that the universe
must cease to exist after finite proper time, which
implies that the universe has spatial topology $S^3$. 
The Second Law of Thermodynamics says the amount of
entropy in the universe cannot decrease, but it can be
shown that the amount of entropy already in the CBR
will eventually contradict the Bekenstein Bound [10]
near the final singularity unless there are no event
horizons, since in the presence of horizons the
Bekenstein Bound implies the universal entropy $S \leq
constant\times R^2$, where $R$ is the radius of the
universe, and general relativity requires $R
\rightarrow 0$ at the final singularity.  The absence of
event horizons by definition means that the universe's
future c-boundary is a single point, call it the {\it
Omega Point}.  MacCallum [11] has shown that an $S^3$
closed universe with a single point future c-boundary
is of measure zero in initial data space.  Barrow [12]
has shown that the evolution of an $S^3$ closed
universe into its final singularity is chaotic.  Yorke
[13] has shown that a chaotic physical system is likely
to evolve into a measure zero state if and only if its
control parameters are intelligently manipulated.  Thus
life ($\equiv$ intelligent computers) almost certainly
must be present {\it arbitrarily close} to the final
singularity in order for the known laws of physics to
be mutually consistent at all times.  Misner [14] has
shown in effect that event horizon elimination requires
an infinite number of distinct manipulations, so an
infinite amount of information must be processed
between now and the final singularity.  The amount of
information stored at any time diverges to
infinity as the Omega Point is approached, since
$S\rightarrow +\infty$ there, implying divergence of
the complexity of the system that must be understood
to be controlled.

\medskip
Let me now expand out the argument in the
preceeding paragraph.  First, let me show in more
detail that unitarity (combined with the Hawking effect
and the Bekenstein Bound) implies that the spatial
topology of the universe must be $S^3$.  The argument I
shall give is independent of the dynamics; it only
depends on the basic structure of quantum mechanics
and Riemannian geometry.  A dynamical argument
would be sticky if one does not want to make any a
priori assumptions about the cosmological constant: a
deterministic (globally hyperbolic) universe with a
negative cosmological constant {\it always} will exist
for only a finite proper time, whatever the spatial
topology [17].  A dynamical proof for $S^3$ can be
found in [18].

\medskip
I have shown in Section 2 that the Bekenstein Bound,
in the presence of particle horizons, implies that each
region of space inside the paricle horizon must have
less than one bit of information when this spatial
region becomes less than a Planck length in radius. 
Since less that one bit physically means that there is
no information whatsoever in the region --- that is, the
laws of physics alone determine the structure of space
--- this region must be isotropic and homogeneous,
because information must be given to specifiy
non-FRW degrees of freedom.  Now the Bekenstein
Bound is not merely a local bound, but a global
constraint, in the sense that it constrains a region
with radius less than the Planck length to have zero
information, rather merely some open ball of with no
apriori minimum size.  But we can overlap these balls
of Planck radius, to conclude that there is no
information anywhere in the spatial universe at the
Planck time.  

\medskip
Now a non-Compact FRW universe at the Planck time
would still be of infinite volume, and thus would
eventually create an infinite number of protons and
neutrons, by the tunnelling process described in
Section 5.  Now Zel'dovich has shown that the lifetime
of a proton to decay via the Hawking process is
$10^{122}$ years (the actual value donesn't matter; it
just has to be finite for my argument).  If the universe
held an infinite number of protons and neutrons, the
probability is one --- a virtual certainty --- that at
least one proton or neutron would decay via the
Hawking process in the next instant after the Planck
time, so the probability is one that unitarity would be
violated.  But unitarity cannot be violated, so the
probability is one that the universe is spatially
compact.

\medskip
We can now apply the Bekenstein Bound to this
comapct universe, and note once again that the
Bekenstein Bound is a global Bound; in particular, 
it implies that the amount of information is zero when
the volume of the universe is the Planck volume.  But
if the universe were not simply connected, the
topology itself would contain information, which is
not allowed.  Hence the universe must be spatially
compact and simply connnected.  The homogeneity and
isotropy of the universe, inferred above, implies that
the spatial sectional curvatures are constant. 
Compactness implies ([16], p. 11) that spatially, the
universe is complete.  It is well-known (e.g., [16], p.
40) that the only complete, simply connected compact
three-manifold with constant sectional curvature is
the three-sphere.

\bigskip
{\bf e. Solution to Standard Model Hierarchy Problem}

\bigskip
Since the validity of the Standard Model of Particle
Physics --- especially of the SM electroweak physics
--- is the essential assumption in this paper, I shall
now further justify this assumption by pointing out
that the standard quantum gravity combined with the
Hawking black hole evaporation effect and the
requirement of unitarity as discussed above,
automatically resolves the Heirarchy Priblem.

\medskip
Recall that the Heirarchy Problem is explaining why
the Higgs mass --- and hence all the particle masses
--- are not dragged up to the the Planck mass (or
higher!) by the self-interactions as expressed by the
renormalization group equation.  Let us first note that
the measurement of the top quark mass at around 175
GeV forces the SM Higgs boson mass to be around 200
GeV, because otherwise the SM Higgs potential would
become unstable due to the higher order quantum
corrections: the highest order term in the Higgs
potential when the quantum corrections are taken into
account is no longer $\lambda\phi^4$, but rather
$C\phi^4\ln(\phi^2/M^2)$ (to one loop order), and the
constant $C$ becomes negative, if the top quark mass
and the Higgs mass becomes greater than about 175
GeV and 200 GeV respectively. (This renormalization
group calculation assumes of course that some
mechanism has already been found to prevent the one
and higher loop self-energy corrections to the mass of
the Higgs boson alone from dragging the Higgs mass to
the Planck mass.)

\medskip
The experimental fact that the SM Higgs vacuum
potential is, given the observed top quark mass, only
marginally stable is of fundamental significance:
when combined with the Hawking effect, it provides
the mechanism that solves the Hierarchy Problem.  

\medskip
Suppose on the contrary that the one and
higher loop corrections to the mass of the Higgs
boson increased the Higgs mass to greater than the
allowed $\sim 200$ GeV.  Then the mass of the Higgs
would be pulled over the vacuum stability bound, and
the mass of the Higgs would grow at least to the
Planck mass, and the mass of the down quark would
thus also increase to within an order of magnitude of
the Planck mass.  But this would mean that a neutron,
with two valence down quarks, would become unstable
via the Zel'dovich effect discussed above to the
formation of a mini-black hole of near Planck mass,
which would then evaporate via the Hawking process,
violating unitarity.  Hence, the one-loop and higher
self-energy terms cannot act to increase the mass of
the Higgs beyond the 200 GeV upper bound allowed by
vacuum stability, since this would violate unitarity.

\medskip
This also shows that the one and higher loop
corrections, which are integrals over the energy in the
loops, necessary have a cut-off at an energy less than
the Planck mass, a cut-off arising from quantum
gravity.  The cut-off is given by the requirement that
the energy in the loop cannot increase to the level that
would result in the formation of a mini-black hole
even virtually.  Thus in spite of naive appearance, this
cut-off is Lorentz and gauge invariant.  To see this,
ignore for a moment all effects except for
self-energy of a single particle.  Then the upper bound
to the value of the energy would be Planck energy,
defined by the condition that no trapped surface of
Planck energy is allowed to be formed, since such
would give rise to a violation of unitarity.  But the
trapped surface condition is a Lorentz and gauge
invariant condition.

\medskip
Notice also that the upper bound to the energy in the
loop integral actually depends on the proper time to
the final singularity.  If the upper bound were the
Planck energy, then the loop correction would give
rise to a Planck-size mini-black hole if the time
before the final singularity were greater than the
Planck time.  Thus, in the earlier period of universal
history --- for example, now --- the cut-off to the
energy allowed in the loop integral must be less than
the Planck energy.  How much would be very difficult
to calculate even given the knowledge of the length of
time before the final singularity.  But the cut-off
must exist, and be less than the Planck energy at the
present time.

\medskip
An extension of this argument allows me to
establish that the quantum gravity theory I have used
in this section, namely the quantum gravity theory
based on the Wheeler-DeWitt equation, is a completly
reliable quantum gravity theory for the early
universe, and in fact the early universe limit of more
general quantum gravity theory that is both
renormalizable and term by term finite.  I refer to the
quantum gravity theory that includes in the Lagrangian
all terms consistent with GL(2,R) symmetry of general
relativity.  It is well-known that if all terms
consistent with this symmetry are included in the
Lagrangian, then gravity is just as renormalizable as
any other theory ([22], p. 506, pp. 517--519; [23],
p.91--92; [24]).  The problem has always been that
there are an infinite number of such terms.  This
objection has been overcome by regarding the resulting
theory as an effective theory, with the higher order
curvature terms coming in only at energies greater
than the Planck energy, but with an apparent
breakdown of the effective theory at energies greater
than the Planck energy.

\medskip
With the Hawking effect and unitarity, we see that no
such breakdown occurs.  Instead, the higher order
curvature terms generate a more intense gravitational
field than the first order Einstein Lagrangian, and thus
would force a mini-black hole at a lower cut-off than
the Einstein term.  This means that in the current
epoch of universal history, these higher order terms
must be completely suppressed by unitarity.  They will
be important near the final singularity --- when the
time before the final singularity is less than the
Planck time --- but they are essentially suppressed
at earlier times, in particular in the present epoch and
near the initial singularity.   So we can ignore these
terms today and in the past, and thus the fact that
adding an infinite number of terms to the Lagrangian
necessarily involves an infinite number of constants
that must be determined by experiment.  The
experiments need not be conducted, indeed cannot be
conducted until the universe is within a Planck time
of the final singularity.  Measuring the values of these
constants are among the infinity of measurements
that must be carried out by life as the universe moves
into the final singularity.  At all times, however, the
``effective'' quantum gravity theory will be term by
term finite, where I have placed the quotes because I
claim that this standard quantum gravity theory can
in fact be regarded as the true theory of quantum
gravity. 

\medskip
Recognizing that the Hawking effect plus unitarity
requires a Lorentz and gauge invariant upper bound to
the energy in a loop integral --- in other words, yields
a Lorentz invariant ultraviolet cut-off --- also solves
the well-known problem of infinite particle
production by time dependent gravitational fields. 
Expressing a quantized field as

$$\phi({\vec x}, t) = (2\pi)^{-3/2}\int
d^3\,{\vec k}[A_{\vec k}\phi_{\vec k}(t)e^{i{\vec
k}\cdot{\vec x}} + A_{\vec k}^\dagger\phi^\ast_{\vec
k}(t)e^{-i{\vec k}\cdot{\vec x}}]$$

The operators $\phi_{\vec k}(t)$ and $\phi^\ast_{\vec
k}(t)$ define what we mean by particles at time $t$. 
Given this definition, the corresponding definition at
time $t_0$ is given by 

$$\phi_{\vec k}(t_0) = \alpha_{\vec k}(t_0)\phi_{\vec
k}(t) + \beta_{\vec k}(t_0)\phi^\ast_{\vec
k}(t)$$

It was established by Parker more than thirty years
ago that the necesary and sufficient condition for the
particle number density to be finite is [19]

$$\int|\beta_{\vec k}(t_0)|^2d^3\,{\vec k} < \infty$$

Since in many cases of physical interest,
$|\beta_{\vec k}(t_0)|^2$ drops off only as $k^{-2}$,
this integral will diverge if the upper limit of the
energy is infinity.  However, the integral is a loop
integral, and thus having the upper bound extend past
the Planck energy would cause the spontaneous
formation of a mini-black hole, which would
immediately evapoate, violating unitarity.  Once again,
this ultraviolet cut-off does not violate Lorentz
invariance, because what is giving  the upper bound is
the non-formation of a trapped surface, and whether a
given 2-sphere is a trapped surface is a Lorentz
invariant condition.  So the definition of particle via
the Hamiltonian diagonalizaton procedure (which is
the definition used above) makes perfect sense given
unitarity and the Hawking effect, so I must disagree
with Fulling who opined in 1979 that no one
should ever again take the idea of Hamiltonian
diagonalization seriously ([19], p. 824). 

\medskip
It has recently been proposed ([26], [27]) that
mini-black holes can be produced at the rate of one per
second in the Large Hadron Collider, due to go on line in
2005.  The above argument shows that this is
absolutely impossible.  No mini-black holes at all will
be produced by the LHC, or by any other accelerator.  My
theory and indeed standard quantum gravity would be
conclusively refuted by the unequivocal observation of
mini-black holes in the LHC.

\medskip
In the above I have made reference only to the down
quarks in the Zel'dovich--Hawking effect.  There
is a reason for omitting the up quarks.  Recall that the
Zel'dovich upper bound is the average time required for
two massive quarks to come within a Schwarzschild
radius of each other, the Schwarzschild mass being
assumed to be the Higgs quark mass.  A particle with
zero Yukawa coupling to the HIggs field would thus
have zero Schwarzschild radius, and thus two such
particles would have an infinite time before coming
within a Schwarzschild radius of each other.   Thus any
massless quark would not be subject to the Ze'dovich
mechanism.  I claim that the mass of the up quark is
probably zero.

\medskip
Recall that the other outstanding theoretical problem
with the Standard Model of particle physics is the 
strong CP problem.  Now that the B factories have seen
CP violation, the solution of spontaneous CP violation
is now ruled out, at least in the sense that all such
models proposed to date predict that CP violation in B
decay should be too small to be observed in the
experiments where it was observed  (I am grateful
to Paul Frampton for a discussion on this point).  The
axion solution is generally considered to be ruled out
by the required fine tuning in the early universe [20]
--- though I would rule it out because the axion has not
been detected.  The only remaining solution to be
strong CP problem is for the mass of up quark to be
zero.

\medskip
Standard current algebra analysis (e.g. [23], p. 231)
giving the ratios of quark masses in terms of the
masses of various mesons indicate that the up quark
has a non-zero mass, but Weinberg ([23], p. 458) points
out that inclusion of terms second order in the strange
quark mass might allow the up quark mass to vanish.
Kaplan and Manohar for example claim [21] that $m_u =
0$ is allowed provided 30\% of the squares of the
meson masses arise from operators second order in the
chiral symmetry breaking, and also that ``The most
striking feature of our result is that a massless up
quark is not in contradiction with the past success of
$SU(3) \times SU(3)$ chiral perturbation theory." ([21],
p. 2006).

\medskip
Setting $m_u =0$ solves the Strong CP Problem, and
including standard quantum gravity effects in the
Standard Model solves the Hierarchy Problem.  Since
these were the main theoretical problems with the
Standard Model, we can be confident in the application
of the Standard Model to the early universe --- and
also confident in the Standard Model's claim that
electromagnetism is not a fundamental field but
instead is a composite of a $U(1)_R$ and an $SU(2)_L$
field.

\medskip
In summary, the one and higher self-energy
corrections to the Higgs boson indeed pull the Higgs
boson mass up to a higher value --- the loop integral
pulls the Higgs (and top quark) mass up to the
maximum value it can have consistent with vacuum
stability.  It cannot pull it up further than this,
because a further value would violate unitarity via the
Hawking effect.  The Hawking effect, by imposing an
upper bound to the energy (ultraviolet cut-off), an
upper bound coming from the requirement that this
stability bound be not exceeded, makes the Standard
Model fully consistent.

\bigskip
{\bf f. Solution to the Cosmological Constant Problem}
\bigskip

I have argued in [25] that the Hawking evaporation
effect plus unitarity prevents the cosmolgical
constant from being exceedingly large, and in fact
requires that the effective cosmological constant, if
ever it becomes small but positive, must eventually
become zero or negative, since otherwise the universe
even if closed would expand forever, resulting in the
evaporation of the black holes which now exist,
violating unitarity.  What I shall now do is describe
the physical mechanism that will eventually
neutralize the observed currently positive effective
cosmological constant.

\medskip
It is well-known that the mutual consistency of the
particle physics Standard Model and general relativity
requires the existence of a very large positive
cosmological constant.  The reason is simple:  the
non-zero vacuum expectation value for the Higgs 
field yields a vacuum energy density of $\sim - 1.0
\times 10^{26}\, {\rm gm/cm^3}(m_H/246)\, {\rm
GeV}$, where $m_H$ is the Higgs boson mass.  Since
this is a negative vacuum energy, it is accompanied by
a positive pressure of equal magnitude, and both the
pressure and energy yield a negative cosmological
constant.  Since the closure density is $1.88\times
10^{-29}\,\Omega_0h^2{\rm gm/cm^3}$, and
observations indicate that  $\Omega_0 = 1$ and $h=
0.66$, there must be a fundamental {\it positive}
cosmological constant to cancel out the negative
cosmological constant coming from the Higgs field. 
What we observe accelerating the universe today is the
sum of the fundamental positive cosmological
constant, and the negative Higgs field cosmological
constant;  this sum is the ``effective'' cosmological
constant.

\medskip
What we would expect is that these two cosmological
constant would exactly cancel, and what must be
explained is why they do not: the vacuum energy
coming from the Higgs field --- more generally, the
 sum of the vacuum energies of all the physical fields
--- is observed to be slightly less in magnitude than
the magnitude of the fundamental positive cosmological
constant.  What must be explained therefore, is why
the vacuum energy sum is slighly less than expected.

\medskip
I shall argue that the instanton tunnelling that has
been shown in Section 5 to generate a net baryon
number also results in the universe being in a false
vacuum slightly above the true vacuum, where, as
expected, the fundamental cosmological constant and
the vacuum energy densities of all the physical fields
do indeed cancel.

\medskip
Recall that the instanton tunnelling works by
non-perturbatively changing the global winding number
of the $SU(2)_L$ field; the winding number is equal to
the number of fermions in the universe.  There is also
a winding number associated with the $SU(3)$ color
force, and the color vacuum --- the
$\theta$-vacuum --- is a weighed sum over all the
winding numbers:  $|\theta> = \sum_n
e^{-in\theta}|n>$.  The fact that $\theta$ is
observed to be essentially  zero is of course the
``strong CP problem'' which I resolved above.

\medskip
There is no necessary connection between the winding
numbers of $SU(2)_L$ and color $SU(3)_L$, but in
fact $\pi_3(G) = Z$ for any compact connected Lie
group $G$, where $\pi_3(G)$ is the third homotopy
group of $G$, expressing that there are non-trivial
mapping of the three-sphere into $G$.  There are thus
{\it three} 3-spheres in cosmology and the Standard
Model:  (1) electroweak $SU(2)_L$ itself, (3) subgroups
of color $SU(3)$ and (3) the spatial 3-sphere.  I
propose that the non-zero winding number due to
mapping of $SU(2)_L$ into itself gives rise to a false
vacuum in one or all of these three, and that the true
vacuum corresponds to a winding number of zero.

\medskip
This means that as long as the number of fermions
minus anti-fermions remains constant on the spatial
3-sphere, the physical fields will remain in the false
vacuum, the effective cosmological constant will
remain positive, and the universe will continue to
accelerate.  Conversely, if instanton tunnelling occurs
in reverse, so that the fermion number of the universe
decreases, then the false vacuum will decrease to the
true vacuum, a state which I have assumed has an
energy density which cancels the positive fundamental
cosmological constant. In the present epoch of
universal history, the winding number remains
constant --- the tunneling probability is very small in
the present epoch --- and thus the sum of the false
vacuum energy density and the fundamental
cosmological constant, this sum being the {\it dark
energy} since it is seen only gravitationally, is
constant.

\medskip
But in the long run, it cannot remain constant, since an
unchanging positive dark energy would cause the
universe to accelerate forever, violating unitarity
when the observed black holes evaporate.  Since the
proton lifetime due to electroweak instanton
tunnelling is greater than the solar mass black hole
lifetime, something must act in the future to speed up
the tunnelling probability.

\medskip
I propose that life itself acts to annihilate protons
and other fermions via induced instanton tunnelling. 
Barrow and I have established that the main source of
energy for information processing in the far future
will be the coversion of the mass of fermions into
energy.  Baryon number conservation prevents this
process from being 100\% efficient, but since the
Standard Model allows baryon non-conservation via
instanton tunnelling, I assume that some means can
and will be found in the far future to allow life to
speed up this process.  So once again the existence of
intelligent life in the far future is required for the
consistency of the laws of physics, since in the
absence of life acting to speed up fermion
annihilation, the universe would accelerate forever,
violating unitarity and incidentally extinguishing life.

\medskip
Since a universe which expanded for a sufficiently
long time would also extinguish life, a universe of the
multiverse which has a radius of maximum expansion
beyond a certain upper bound cannot develop
structure:  such structure would necessarily mean
that the entropy is non-zero, and in Section 2, I showed,
following Bekenstein, that in the presence of event or
particle horizons, the entropy of the universe has to
approach zero near a singulairty, and only life can
force the elimination of horizons.  So for universes
which have a radius at maximum expansion greater than
this upper bound, both the initial and final
singularities are Friedmann, with the entropy
remaining zero for all of these very large universe's
history.   Similarly, universes which have a radius
sufficiently small can never develop structure, foe
life will never have time to evolve.  (Universes whose
radius at maximum expansion is less than the Planck
length never develop structure, because the
Bekenstein Bound never allows the generation of
information at all; the entropy starts and remains
zero from the Bekenstein Bound alone.)

\medskip
So there is a narrow band of universes in the universe
wherein entropy, structure, and life mutually exist.  
In these universes, the initial singularity is
Friedmann, with zero entropy sufficiently close to the
singularity, and with entropy that diverges as the
final singularity is approached.  This implies the
Penrose condition on the initial and final singularity:
an initial singularity is dominated by the Ricci
curvature (Friedmann singularity) and a final
singularity is dominated by Weyl curvature ---
dominate Weyl curvature is a necessary feature of
Mixmaster oscillations which are required to
eliminate event horizons.

\medskip
The multiverse with the narrow band of universes
containing entropy structure and life is pictured in
Figure 6.1.

\bigskip
Figure 6.1: The Multiverse, and the entropy, structure
forming and life band.  For universes in the band, the
initial singularity is isotropic and homogeneous with
zero entropy, while the final singularity has infinite
entropy, structure and Weyl curvature.  Universes
outside the band start and remain at zero entropy, never
developing structure, Weyl curvature, or life.

\bigskip
\centerline{\bf References}
\bigskip

\noindent\hangindent=20pt\hangafter=1
\item{[1]} Frank  J. Tipler 1986.  Physics Reports {\bf
137}: 231.

\medskip
\noindent\hangindent=20pt\hangafter=1
\item{[2]}  V.G. Lapchinskii and  V.A. Rubakov 1977. 
Theoretical and Mathematical Physics {\bf 3} 1076.

\medskip
\noindent\hangindent=20pt\hangafter=1
\item{[3]} Bernard F. Schutz 1971.   Physical Review
{\bf D4}: 3559.

\medskip
\noindent\hangindent=20pt\hangafter=1
\item{[4]} Max Jammer, {\it The Philosophy of Quantum
Mechanics}, (New York: Wiley 1974).

\medskip
\noindent\hangindent=20pt\hangafter=1
\item{[5]}  N. Bohr, ``Discussion with Einstein on
Epistemological Problems in Atomic Physics,'' in {\it
Albert Einstein: Philosopher-Scientist}, edited by P.
A. Schilpp (New York: Harper Torchbooks 1959).

\medskip
\noindent\hangindent=20pt\hangafter=1
\item{[6]} Lev D. Landau and E. M Lifshitz, {\it Quantum
Mechanics: Non-relativistic Theory, 3rd edition}
(Oxford: Pergamon Press, 1977).

\medskip
\noindent\hangindent=20pt\hangafter=1
\item{[7]} David Bohm, Phys. Rev. {\bf 85} (1952),
166 and 180; Phys. Rev. {\bf 89} (1953), 458.

\medskip
\noindent\hangindent=20pt\hangafter=1
\item{[8]} Arno B\"ohm {\it The Rigged Hilbert Space
and Quantum Mechanics,} Lecture Noties in
Physics no. 78. (Berlin: Springer-Verlag, 1978); second
edition (with M. Gadella) {\it Dirac Kets, Gamow
Vectors, and Gel'fand Triplets,} Lecture Noties in
Physics no. 348. (Berlin: Springer-Verlag, 1989).

\medskip
\noindent\hangindent=20pt\hangafter=1
\item{[9]} S.W. Hawking, ``Breakdown of Predictability
in Gravitational Collapse,'' Physical Review {\bf D14 }
(1976), 2460.

\medskip
\noindent\hangindent=20pt\hangafter=1
\item{[10]} Jacob D. Bekenstein, ``Is the
Cosmological Singularity
Thermodynamically Possible?'' International Journal of
Theoretical Physics {\bf 28} (1989), 967.

\medskip
\noindent\hangindent=20pt\hangafter=1
\item{[11]} Malcolm MacCallum

\medskip
\noindent\hangindent=20pt\hangafter=1
\item{[12]} John D. Barrow, Physics Reports {\bf 85}
(1982), 1.

\medskip
\noindent\hangindent=20pt\hangafter=1
\item{[13]} J. A. Yorke {\it et al} Physical Review
Letters {\bf 65} (1990), 3215; Physical Review
Letters {\bf 68} (1992), 2863.

\medskip
\noindent\hangindent=20pt\hangafter=1
\item{[14]} Charles W. Misner, Physical Review
Letters {\bf 22} (1969), 779; Astrophysical Journal
{\bf 151} (1968), 431. 

\medskip
\noindent\hangindent=20pt\hangafter=1
\item{[15]} Robert M. Wald, {\it Quantum Field Theory
in Curved Spacetime}, (Chicago: Chicago University
Press, 1994).

\medskip
\noindent\hangindent=20pt\hangafter=1
\item{[16]} Jeff Cheeger and David G. Ebin, {\it
Comparision Theorems in Riemannian Geometry},
(Amsterdam: North-Holland, 1975)

\medskip
\noindent\hangindent=20pt\hangafter=1
\item{[17]} Frank J. Tipler, ``Negative Cosmological
Constant'', Astrophysical Journal 1975.

\medskip
\noindent\hangindent=20pt\hangafter=1
\item{[18]} Barrow, Galloway, and Tipler, Monthly
Notices of the Royal Astronomical Society.

\medskip
\noindent\hangindent=20pt\hangafter=1
\item{[19]} Steve A. Fulling, ``Remarks on Postive
Frequency and Hamiltonians in Expanding Universes,''
Gen. Rel. and Grav. {\bf 10} (1979), 807--824.

\medskip
\noindent\hangindent=20pt\hangafter=1
\item{[20]} Holman et al PL B282, 132 (1992).

\medskip
\noindent\hangindent=20pt\hangafter=1
\item{[21]} David B. Kaplan and Aneesh V. Manohar,
 ``Current-Mass Ratios of the Light Quarks'', Phys.
Rev. Lett. {\bf 56} (1986), 2004.

\medskip
\noindent\hangindent=20pt\hangafter=1
\item{[22]} S. Weinberg, {\it The Quantum Theory of
Fields, Volume I} (Cambridge University Press,
Cambridge, 1996)

\medskip
\noindent\hangindent=20pt\hangafter=1
\item{[23]} S. Weinberg, {\it The Quantum Theory of
Fields, Volume II} (Cambridge University Press,
Cambridge, 1996).

\medskip
\noindent\hangindent=20pt\hangafter=1
\item{[24]} John F. Donoghue, ``General Relativity as an
Effective Field Theory: The Leading Quantum
Corrections,'' Phys. Rev. {\bf D50} (1994), 3874--3888.

\medskip
\noindent\hangindent=20pt\hangafter=1
\item{[25]} Frank J. Tipler, ``The Ultimate Fate of the
Universe, Black Hole Event Horizons, Holography and
the Value of the Cosmological Constant,''
astro-ph/0104011.

\medskip
\noindent\hangindent=20pt\hangafter=1
\item{[26]} S. Dimopoulos and G. Landsberg, ``Black
Holes at the Large Hadron Collider,'' Phys. Rev. Lett.
{\bf 87} (2001), 161602.

\medskip
\noindent\hangindent=20pt\hangafter=1
\item{[27]} S. B. Giddings and S. Thomas, ``High Energy
Colliders as Black Hole Factories: the End of Short
Distance Physics.'' Phys. Rev D (in press).

\vfill\eject

\font\bigtenrm=cmr10 scaled\magstep5
\centerline{\bf{\bigtenrm 7. The SU(2) Gauge Field
and the}}
 \bigskip
\centerline{\bf{\bigtenrm Higgs Field in the Present
Day Epoch}}

\bigskip
\bigskip

\centerline{\bf a.Dynamics of the Pure SU(2) and
Higgs Field}
\bigskip
\bigskip

The Bekenstein Bound requires the Higgs field to
intially have the value $\phi =0$, as has been
discussed in the privious section.  Since the present
day value is $<\phi> = 246\, {\rm GeV}$, the Higgs
field must have rolled down the incline of its
potential.  It is usual to assume that this Higgs energy
has long since been thermalized into the CMR, but as I
have shown, the CMR gauge field componet would have a
Planck distribution in a FRW background whether it is
thermal or non-thermal, I shall instead investigate the
possibility that the Higgs energy has never been
thermalized, but instead has been partly absorbed by
the SU(2) gauge field, and partly decreased in density
by the expansion of the universe.

\medskip
Let us recall what the expansion of the universe must
have been like so that the nucleosyntheis calculations
are still vaild, and so that the structure formation 
computer simulations are also still valid.  During the
nucleosynthesis era, the expansion rate must have
been $R(t)\sim t^{1/2}$ corresponding to radiation
domination --- as discussed earlier, any massless
gauge field will generate this behaviour if it is the
dominant energy density in the universe.  After
decoupling, the expansion rate must change from the
radition domination rate to the matter dominated rate
of $R(t) \sim t^{2/3}$ with a slight mixture of a
$\Lambda$ term, the change being required by
structure formation simulations, which agree best
with the $\Lambda{\rm CDM}$ model (actually, the
best results are obtained by assuming that the CDM is
slightly ``warm'', but as we shall see, this will be
allowed in my proposal for the dark matter).  The
nucleosynthesis data indicate that baryons are only a
small fraction of the dark matter --- which by
definition is that ``substance'' which is responsible for
the $R(t) \sim t^{2/3}$ expansion rate.

\medskip
In my model there are only two fields outside of the
baryons: the SU(2) gauge field and the Standard Model
Higgs field.  I shall now argue --- but not prove ---
that it is {\it possible} for these two fields
interacting together to produce the observation CMBR
and the dark matter.  (I have shown in Section 6 how
the Standard Model vacuum naturally provides a
non-zero cosmological constant; this I will take to be
the dar energy).

\medskip
Let us first consider the time evolution of the Higgs
field by itself, and then consider its interaction with
the SU(2) gauge field.  For a general scalar field
Lagrange density of the form $-{1\over2}\partial_\mu
\phi\partial^\mu\phi - V(\phi)$, the equation of
motion for the scalar field will be

$${d\over dt}({1\over2}\dot\phi^2 + V) =
-3H\dot\phi^2$$

\noindent
or

$$d\rho/dt = -3H(\rho + p) \eqno(7.1)$$

\noindent
where
$$\rho = \dot\phi^2 + V(\phi)\eqno(7.2)$$

\noindent
is the energy density of the scalar field, and 

$$p = \dot\phi^2 - V(\phi)\eqno(7.3)$$

\noindent
is the pressure of the scalar field.

\bigskip
\bigskip
\centerline{\bf b. SOLUTION TO ``DARK MATTER''
PROBLEM:} 
\medskip
\centerline{\bf WHAT IT IS AND HOW IT HAS ELUDED
DETECTION}
\bigskip
\bigskip

Turner has shown [3] that if we regard the scalar field
as the sum of a rapidly oscillating part, and a slowly
varing part, then a scalar potential of the form
$V(\phi) = a\phi^2$, which is the approximate form of
the SM Higgs potential in the present epoch, would
give rise to an effective mass density that would drop
of as $R^{-3}$, just as pressureless dust would.  I
conjecture that the combination of the SM Higgs field
coupled to a pure $SU(2)_L$ field would naturally
split into two fields that would appear to evolve
independently, one dropping off as $R^{-3}$, and the
other dropping off as $R^{-4}$.  One would be the CMBR,
and the other would be the dark matter.  Recall that
the Z boson has all the quantum numbers as a photon,
and in fact can be made to form superpositions with
photons.  The interaction strength of the Z with
fermions is stronger than the photon, and the only
reason that the Z boson acts weakly is its large mass.  
Similarly, the candidate I am proposing as the dark
matter will interact only weakly with fermions
because it is basically a Z particle.

\medskip
If this conjecture is correct, then the reason the dark
matter has not been detected is because it must
necessarily always be found in accompanied with
$SU(2)_L$ pseudo-photons, and all the experiments to
detect the dark matter have carefully been designed to
eliminate all photon interactions.

\medskip
Of course, the reason why such a possible dark matter
candidate has heretofore not been considered is that
it has been thought that the rapid oscillations of a SM
Higgs field would quickly decay away ([3], section IV;
[4]), into photons.  I would conjecture that this is
indeed what happens; the Higgs field decays into the
$SU(2)_L$ field, which then passes the energy back
into the Higgs field.

\medskip
Let me emphasize (as if it needed emphasizing) that
these are very counter-intuitive conjectures I am
making, and I have given no mathematical evidence
that the combined Higgs coupled to a pure $SU(2)_L$
field could in fact behave this way.  I instead can only
offer an {\it experimental} argument that something
like this scenario must be in operation:  it has been
known for 35 years that ultra high energy cosmic rays
propagate through the CMBR as if the CMBR were not
present, and as I shall demonstrate in Section 9, this
is possible if --- and if the SM is true, only if ---
the CMBR has the properties of a pure $SU(2)_L$ field. 
And we have no laboratory experimental evidence that
the SM is incorrect.  The SM has passed every test we
have put it through for the past 25 years.

\bigskip
\bigskip
\centerline{\bf c. WHY AN SU(2) COMPONENT WOULD
HAVE}
\medskip
 \centerline{\bf NO EFFECT ON EARLY UNIVERSE
NUCLEOSYNTHESIS}
\bigskip
\bigskip
The baryons, once created by the mechanism in section
5, would be in a Planck distribution gauge field, with
thermal properties identical to the usual standard
cosmological model.  Recall that the interaction
constants of the charged particles with the radiation
field are relevant only to force the particles to also be
in a thermal distribution like the radiation field. 
Thus, the reduced interaction strength of a pure
$SU(2)_L$ field (discussed at length in Sections 8 and
9) would have no effect on the distribution of the
particles and thus on nucleosynthesis.  (The same
would be true of the fluctuation spectrum observed in
the acoustic peaks.  As mentioned in Section 6,
flatness requires a Harrision-Zel'dovich spectrum for
the fluctuations, and the magnitude of the fluctuation
spectrum is fixed by the requirement that the
fluctuations be due entirely from the creation of
baryons.

\bigskip
\bigskip
{\bf d. SUPPRESSING EARLY UNIVERSE PAIR CREATION,
INVERSE AND DOUBLE COMPTON, and THERMAL
BREMSSTRAHLUNG}
\bigskip
\bigskip

I have argued in previous Sections that in the
beginning, the universe must have contained nothing
but a pure $SU(2)_L$ field.  Even if this were true, one
might think that this pure $SU(2)_L$ field would have
long before the de-coupling time around a redshift of
1,000, this pure state would have thermalized into a
normal EM field.  I cannot prove that there is {\it no}
mechanism that would have resulted in the
thermalization of the proposed pure $SU(2)_L$ field,
but I can demonstrate that the standard ([5], [6], [7])
three main mechanisms of thermalization in early
universe cosmology, namely pair creation, double
compton scattering, and thermal bremsstrahlung
actually will not thermalize a pure $SU(2)_L$ field.

\medskip
An outline of the proof is simply to write down the
Feynman Diagrams for all three processes, (actually
only two; the Diagram for pair creation is essentially
the same as the Diagram for Bremsstrahlung), and
realize that each ``pseudo-photon'' of the pure
$SU(2)_L$ field can couple {\it only} to left-handed
electrons, and right-handed positrons.  It is quickly
noticed that the Diagrams violate conservation of
angular moemntum; all of these processes require a
spin flip involving the same particle, and this is
impossible.  The no-go theorem is in all essentials the
same as well-known decay asymmetry of the W boson.

\bigskip
\centerline{\bf References}
\bigskip

\noindent\hangindent=20pt\hangafter=1
\item{[1]}  John D. Barrow and Paul Parsons,
``Inflationary Models with Logarithmic Potentials,''
Phys. Rev. {\bf D52} (1995), 5576--5587.

\medskip
\noindent\hangindent=20pt\hangafter=1
\item{[2]} George F. R. Ellis and Mark S. Madsen, ``Exact
Scalar Field Cosmologies,'' Class. Quantum Grav. {\bf
8} (1991) 667--676.

\medskip
\noindent\hangindent=20pt\hangafter=1
\item{[3]} Michael S. Turner, ``Coherent Scalar Field
Oscillations in an Expanding Universe,'' Phys. Rev. {\bf
D28} (1983, 1243--1247.

\medskip
\noindent\hangindent=20pt\hangafter=1
\item{[4]} Sean Carroll, Phys. Rev. Lett.

\medskip
\noindent\hangindent=20pt\hangafter=1
\item{[5]} Alan P. Lightman, ``Double Compton Emission
in Radiation Dominated Thermal Plasmas,'' Ap. J. {\bf
244} (1981), 392--405

\medskip
\noindent\hangindent=20pt\hangafter=1
\item{[6]} L. Danese and G. De Zotti, ``Double Compton
Process and the Spectrum of the Microwave
Background,'' Astro. Astrophys. {\bf 107} (1982),
39--42.

\medskip
\noindent\hangindent=20pt\hangafter=1
\item{[7]} C. Burigana, G. De Zotti, and L. Danese,
``Constraints on the Thermal History of the Universe
From the Microwave Background Spectrum,'' Ap. J. {\bf
379} (1991), 1-5.

\vfill\eject

\font\bigtenrm=cmr10 scaled\magstep5
\centerline{\bf{\bigtenrm 8. Detecting an SU(2)
Component}}
 \bigskip
\centerline{\bf{\bigtenrm In the Cosmic Microwave
Background}}
 \bigskip
\centerline{\bf{\bigtenrm With the Original CMBR
Detectors}}
 \bigskip
\centerline{\bf{\bigtenrm And Using a Penning Trap}}

\bigskip
\bigskip

{\bf a. Right-handed electrons Won't Couple to an SU(2)
CBR Component} 
\settabs 9 \columns

\bigskip
Anyone contemplating a CBMR experiment should first
familarize him/herself with the basic experimental
techniques.  These are described in detail in Bruce
Partridge's excellent book [3].  The experiments
described in this section will be rahter
minor modifications of the basic CMBR experiments.

\medskip
The main effect of the CBMR being a pure (or as we shall
see, almost pure) $SU(2)_L$ gauge field is that in this
case, {\bf the CMBR will not couple to right-handed
(positive helicity) electrons}, while standard
electomagnetic radiation couples to electrons of both
helicities with equal strength.  All the experimental
tests of the almost pure $SU(2)_L$ hypothesis which I
shall propose in this section are based on this crucial
property.  But before reviewing the experimental
tests, let me first discuss the question of coupling
strength of the left-handed electrons with a CMBR
which is pure $SU(2)_L$.

\medskip
Recall that in the Standard Model, the $U(1)_R$ gauge
field playes three roles.  First and formost, it allows
the EM field to couple to right-handed electrons. 
Second, it forces a distinction between the $Z^\mu$
gauge field and the EM field 4-potential $A^\mu$.
Finally, it allows the unit normalizations of the
$U(1)_R$ and the $SU(2)_L$ fundamental gauge fields
$B^\mu$ and $W_j^\mu$ respectively to be carried over
to the physical gauge fields $Z^\mu$ and $A^\mu$. 
These latter two properties are usually termed the
``orthogonality" and ``normality" properties.  The
orthogonality and normality properties are at risk
when there is no $U(1)_R$ gauge field at all, so I shall
propose that the actual CMBR contains a small
admixture of $U(1)_R$ to maintain these key
properties.  I would expect the energy density of the
$U(1)_R$ component to be of the order of the energy
density of the anisotropic perturbations in the CMBR,
which would be the source of the small $U(1)_R$
component (recall that in the very early universe, the
radiation field which is the sole matter constituant of
the universe {\it must} be pure $SU(2)_L$).

\medskip
In the Standard Model the gauge fields are related by

$$A^\mu = {{g_2B^\mu + g_1W_3^\mu}\over \sqrt{g_1^2
+ g_2^2}}\eqno(8.1)$$

$$Z^\mu = {{-g_1B^\mu + g_2W^\mu_3}\over
\sqrt{g_1^2 + g_2^2}}\eqno(8.2)$$

\noindent
where $g_1$ and $g_2$ are respectively the $U(1)_R$
and the $SU(2)_L$ gauge coupling constants.  It is
clear from $(8.1)$ and $(8.2)$ that if the fundamental
fields $B^\mu$ are normalized to unity, then so are
$A^\mu$ and $Z^\mu$, and also that the latter two
fields are orthogonal if the former two are orthognal. 
It is also clear that the real reason for the
normalizatoins is to force the the EM fieldt to couple
with equal strength to both left and right handed
electrons.  But it is this equality that I am proposing
does not exist in the case of the CMBR.

\medskip
The coupling to electrons in the SM Lagrangian is

$${\bar e_L}\gamma_\mu e_L\left[ {g_1\over2}B^\mu
+ {g_2\over2}W_3^\mu\right] + {\bar
e_R}\gamma_\mu e_Rg_1B^\mu\eqno(8.3)$$

Suppose now that we set $B^\mu =0$.  Solving $(8.1)$
for $W^\mu_3 = (\sqrt{g_1^2 + g_2^2}/g_1)A^\mu$  ---
in which case a normalized $W_3^\mu$ does {\it not}
yield a normalized $A^\mu$ and substituting this into
$(8.3)$ gives

$$e_{EM}\left(\alpha_2\over2\alpha\right)A^\mu{\bar
e_L}\gamma_\mu e_l\eqno(8.4)$$

\noindent
where $\alpha_2 = 1/32$ is the $SU(2)_L$ fine
sturcture constant, and $\alpha = 1/137$ is the usual
fine structure constant.  So if we accept the
normalizaton of $(8.3)$, the coupling between
electrons and the pure $SU(2)_L$ field would be
increased relative to the usual $e_{EM}$.  However, I
would argue that either a small admixture of $B^\mu$
would force the usual coupling between the CBMR even
if mainly $SU(2)_L$, or else the appropriate
normalizaton to use in computing the interaction
between a CMBR which is almost pure $SU(2)_L$ is to
normalize $A^\mu$ even if it is mainly pure
$SU(2)_L$.  But I've gone through this calculation to
point out that there {\it may} be a different coupling
between a CMBR that is almost pure $SU(2)_L$, and the
usual $A^\mu$ field CMBR.  I doubt this possibility,
because a stronger coupling would ruin the early
universe nucleosyntheis results.  The stronger
coupling would also ruin the ultrahigh energy cosmic
ray effect which I shall discuss in Section 9.

\bigskip
\bigskip
{\bf b. Detecting an SU(2) Component Using Hans
Dehmelt's Penning Trap} 
\bigskip
\bigskip

Hans Dehmelt's Penning Trap ([5], [6]) is the ideal
instrument to test the idea that the CMBR will not
interact with right-handed electrons.  The basic
structure of the Penning Trap is pictured in Figure 8.1.

\bigskip
\centerline{Figure 8.1: the Penning Trap}
\medskip
 (Figure 1 on page 17 of Dehmelt's Am. J. Phys.
article). 

\bigskip
Figure caption: Penning Trap (Taken from
Dehmelt [6]) The electron orbit is a combination of
vertical motion due to the electric field (pictured), and
a circular cyclotron motion due the the magnetic field
$\vec B_0$.  The pictured assembly is placed in an
ultrahigh vacuum, and cooled to liquid helium
temperature $\sim 4$ K.  Cap to cap separation is
about $0.8$ cm.  

\medskip
In a Penning Trap, a single electron (or positron) is
captured by a vertical magnetic field, and an electric
field due to charges on a curved ring and two caps.  In
the Seattle Penning Trap, cap to cap separation is
about $0.8$ cm, the magnetic field $\vec B_0$ was 5
T.  The magnetic field results in a circular cyclotron
motion at $\nu_c = e{\vec B_0}/2\pi m_e = 141\, {\rm
GHz}$, where $e$ and $m_e$ are the electron charge
and electron mass respectively. The charge on the
ring and the caps is adjusted to give a weak
quadrupole field with potential well depth $D= 5\,
{\rm eV}$, yielding an axial oscillation frequency of 64
MHz. (The electron feels a simple harmonic restoring
force with spring constant $k = 2D/Z_0^2$), where
$2Z_0$ is the cap to cap separation.

\medskip
If two turns of nickel wire are wrapped around the ring
electrode, the large applied magnetic field magnetizes
this, and this ``bottle field'' interacts with the
electron's magnetic moment, allowing the spin of the
electron to be continuously measured.  This
``continuous Stern-Gerlach effect'' forces the electron
to be in one of its spin states, and it is possible to
determine which one the electron is in, and to measure
transitions between the two spin states.

\medskip
The energy  of the cyclotron motion of the electron is
quantized, with energy

$$E_n = (n+{1\over2})\,h\nu_c$$

At 4 K, observations give $<n> \approx 0.23$, and for
intervals of about 5 second, the electron is observed
in the $m=-1/2$ state or the $m=+1/2$ state ([5], p.
543).  With $\nu_z = 64\,{\rm MHz}$, this means that if
the state is chosen to be the $m=-1/2$, the electron
will have positive helicity (be right-handed) for
one-half the time for 128 million cycles --- while the
electron is moving down, and negative helicity (be
left-handed) for the other half of the time; that is,
when the electron is moving up. 

\medskip
The electron can undergo a spin flip $\Delta n =0, \,m =
+1/2 \rightarrow +1/2$. This is actually the result
of two separate transitions:  the transition
$n= 0 \rightarrow 1$ is induced by the 4 K thermal
radiation, and transition is followed by the transition
$(n=1,\, m=-1/2) \rightarrow (n=0,\, m=+1/2)$ induced
by an applied rf field ([5], p. 543).

\medskip
The key point to note that the thermal transition,
were the electron with $m=-1/2$to be immersed in the
CMBR thermal field and were the CMBR to be a pure
$SU(2)_L$ field, the thermal transition could occur
only one-half of the  time, that is, when it is
moving up, when it it has left-handed helicity.  That
is, {\bf the ``thermal'' transition rate in a pure
$SU(2)_L$ field would be one-half the transition rate
in a pure electromagnetic heat bath}.  Thus the
Penning Trap can be used to determine whether the
CMBR is indeed pure EM, or instead pure $SU(2)_L$, as
I am claiming.  An experiment to test this would
introduce CMBR radiation via the gap between the ring
electrode and the cap electrodes.

\medskip
In the actual experiment, of course, the radiation
from the cap and ring electrodes would in fact be
thermal EM radiation, and this would induce
transitions at all times in the electron's motion.  If
there were no such radiation, the transition rate
would increase from an average of 5 seconds to 10
seconds, but the actual transition rate would be
proprotional to the ratio of area between the
electrodes to the area of the electrodes that face the
cavity where the electron moves.

\medskip
From the drawing of the Penning Trap in Figure 8.1,
one might infer that this ratio would be quite small,
but appearances can be deceiving.  More precise
drawings of the Penning Trap can be found in ([7], p.
235; [9], p. 108).  I reproduce the more precise drawing
from [7] below as Figure 8.2.

\bigskip
\centerline{Figure 8.2: Scale Drawing of Penning Trap}
\bigskip

In effect the electron is in the center of a spherical
region whose center is the Penning Trap pictured in
Figure 8.2.  Let us approximate the area ratio as
follows.  Take a sphere of radius $a$, and intersect it
with two coaxial cylinders, with the axis of both
passing through the center of the sphere.  Let the radii
of the two cylinders be $r_{in}$ and $r_{out}$, with
$r_{in} < r_{out} < a$.  Then the area of the sphere
between the cylinders is

$$A = 4\pi a^2\left[{ 1 + \sqrt{1-\left(r_{out}\over
a\right)^2} - \sqrt{1-\left(r_{in}\over
a\right)^2}}\right]$$

This area is a good approximation to the gap between
the ring electrode and the cap electrode.  If we feed
the signal from the CMBR thorough only the gap
between the upper cap and the ring electrode, then the
available singla area would be $1/2$ the above area. 
Making a rough measurement of the figure in ([9], p.
108), I obtained $A/4\pi a^2 = 0.49$, and if this is
accurate, as much as $1/4$ of the ``thermal'' radiation
inducing a state transition can be a signal from outside
the Penning Trap, assuming only the upper gap is used
(as the terminus of a circular wave guide).

\medskip
In other words, the outside signal will in effect be
transmitted through a coaxial wave guide, for which
the gap between the upper cap and the ring electrode is
the terminus.  Recall that a coaxial wave guide can
transmit TE and TM, as well as TEM waves.  The power
flow through a coaxial wave guide is calculated by
all physics graduate students([1], p. 385) to be

$$P = \left[c\over 4\pi\right]\sqrt{\mu\over\epsilon}
\pi r_{in}^2|H_{in}|^2\ln\left(r_{out}\over r_{in}
\right) = \left[c\over
4\pi\right]\sqrt{\mu\over\epsilon}
\pi r_{out}^2|H_{out}|^2\ln\left(r_{out}\over r_{in}
\right)$$

\noindent
Or, if $|H| = |E|$, as it will be for TEM waves, and we
assume Gaussian units (in which case the factor in
brackets comes into play and $\mu = \epsilon = 1$
for a vacuum or air wave guide), the power passing
through the wave guide will be

$$P = c\rho_{in}\pi r^2_{in}\ln\left(r_{in}\over
r_{out}\right) \approx 2c\rho_{av}A$$

\noindent
where $\rho_{in}$, $\rho_{av}$, and $A$ are the energy
density at the inner radius, the average energy density
in the annulus, and the area of the open annulus
respectively, and I have assumed that $(r_{out} -
r_{in})/r_{in} << 1$.  So the power flow from the
out side is just the flow on would expect through an
opening of the size of the gap; the walls have no
significant effect.

\medskip
Of course, the signal from outside the Penning Trap
will consist of radiation from several sources, only
one of which will be CMBR.  The other radiation
sources will be EM field radiation, and will couple to
right-handed electrons.  The various other sources are
discussed at length in ([3], pp. 103--139), and a
shorter introduction to these other sources can be
found in the original papers, e.g. ([10], [11]). 
Remarkably, the main non-CMBR is 300 K radiation
from the ground, and if the detector is shielded from
this radiation --- easy to do with metal reflectors
preventing the ground radiation from reaching the
detector antenna --- then the other radiation sources
all have a radiation temperature of the order of a few
degrees K, and there are methods to measure them
independently, and thus subtract them out.

\medskip
For example, the atmosphere has a zenith temperature
of about 4 K, but as the optical depth depends on the
angle $z$ from the zenith as $\sec z$, the atmosphere
temperature goes as $T_{atm}(z) = T_{atm}(0)\sec z$,
and thus by making a series of measurements of the
total energy received by the antenna at several angles
$z$, the energy of the atmosphere can be subtracted
out (details [3], pp 120--121).

\medskip
Since the transtion probability $(n=0) \rightarrow
n=1)$ depends on the square of the cyclotron
frequency [7], the transition rate due to the CMBR will
be too low unless the frequency looked at is near the
5 T cyclotron frequency  $\nu_c = 141\, {\rm GHz}$. 
This is much higher than the window of 3 to 10 GHz
used in the classical CMBR measurements.  However,
there is an atmospheric window at 0.33 cm, or 91 GHz,
sufficiently near the 5 T cyclotron frequency that the
transition rate would be reduced only by a factor of
$(91/141)^2 = 0.42$, and the required 3.2 T Penning
Trap magnetic field should be easy to achieve.   The
CMBR observation at 91 GHz, however, is best
conducted at high altitudes (the first CMBR
measurement was conducted at the High Altitude
Observatory at Climax Colorado which was at an
altitude of $11,300, {\rm ft}$.   The instrument was
actually tested at Princeton University, where it was
built, but even in the winter, the water vapor at
Princeton made the measurement of the $\sec z$
atmosphere contribution difficult to eliminate (D.T.
Wilkinson, private communication).  But in principle,
the 91 GHz CMBR measurement could be done (though
with difficulty) even at Seattle or Cambridge, MA,
where Penning Traps are in operation.  It would better
done with the operational Penning Trap at Boulder CO,
which is already art a high altitude, and the water
vapor is naturally low.

\medskip
Although I have emphasized that the first effect one
should search for with the Penning Trap is the
reduction in transition rate due to the fact that the
CMBR can interact with the Penning Trap electron only
for 1/2 the time, an even better test would be to
observe that the transition occurs only in that part of
the electron's orbit when the electron is left-handed,
for example when a spin down electron is moving up,
and when a spin up electron is moving down.  With
positrons, the situation would be reversed: since the
$SU(2)_L$ field can couple only to right-handed
positrons, a spin up positron should be able to
interact with a pure $SU(2)_L$ CMBR only when the
positron is moving up, and a spin down positron
would be able to interact only when it was moving
down.  However, such a measurement would be
difficult given the standard voltage between the cap
and ring electrodes, which yield the 64 MHz vertical
motion frequency.

\bigskip
\bigskip
{\bf c. Detecting an SU(2) Component With the Original
CMBR Detectors with Filters}

\bigskip
\bigskip
As I said above, the Penning Trap is the ideal
instrument to determine whether or not the CMBR is
indeed a pure $SU(2)_L$ field, or simply an EM field. 
Unfortunately, setting up a Penning Trap to look for
the expected difference is quite an expensive
proposition; a series of e-mails between myself and
Hans Dehmelt's group indicated that it would take \$
250,000 and more to set up such an experiment, to say
nothing of the difficulty of moving the instrument to
the best location, a dry desert high altitude plateau. 
For this reason, it would be nice if a quick and
inexpensive test of the pure $SU(2)_L$ hypothesis
could be found.  In this subsection, I shall outline such
an experiment, but I should caution the reader that the
proposed apparatus depends on estimates on the
behaviour of electrons in conductors and
semi-conductors when an $SU(2)_L$ field interacts
with electrons in such a material, and these estimates
might not be reliable.  So a null result {\it might} not
rule out the pure
$SU(2)_L$ hypothesis.  On the plus side, a positive
result {\it would} confirm the $SU(2)_L$ hypothesis,
and the quick and dirty experiment I shall now propose
can be done with a simple modification of the original
apparatus set up by the Princeton group in 1965 to
detect the CMBR.  Even if the original apparatus no
longer exists, it can be re-built for the cost of at
most a few thousand dollars, and a single day's
observation should suffice to see if the effect is
present.

\medskip
The basic physical effect I shall use is as
follows.  Since a pure $SU(2)_L$ CMBR field will not
couple to electrons with positive helicity, a CMBR
wave will penetrate deeper into a conductor than an
ordinary EM wave, since in a conductor at room
temperature the conduction electrons have on the
average zero net helicity: half on the average have
positive helicity and the other half have negative
helicity.  I shall now show how to use this fact to
confirm that the CMBR is a pure $SU(2)_L$ field.

\medskip
The transmission coefficient for EM waves into an
infinite conducting plate with vacuum on either side
has been calcuated by Stratton ([15], pp. 511--516).  I
shall repeat his derivation because it will allow me to
point out some of the problems that may arise using
the filter experiment rather than the Penning Trap to
detect a pure $SU(2)_L$ CMBR.  

\medskip
Let us, following Stratton, imagine that we have three
arbitrary homogeneous media labeled (1), (2), and (3),
with dielectric constants $\epsilon_1$, $\epsilon_2$,
$\epsilon_3$, magnetic permeabilities $\mu_1$,
$\mu_1$, $\mu_1$, and propagation factors $k_1$,
$k_2$, and $k_3$ respectively.  The thickness of the
intermediate medium (2) will be $d$.  In medium (1), we
only have magnitudes of the incident and reflected
waves:

$$E_i = E_0e^{ik_1x - i\omega t}, \,\,\,\,\, H_i =
{k_1\over \omega\mu_1}E_i$$

$$E_r = E_1e^{-ik_1 x - i\omega t}, \,\,\,\,\, H_r =
-{k_1\over \omega\mu_1}E_r$$

The EM field in the middle medium (2) will contain
wave which are moving to the right and waves which
are moving to the left:

$$E_m = (E^+_2e^{ik_2x} + E^-_2e^{-ik_2x})e^{-i\omega
t}$$

$$H_m = {k_2\over \omega \mu_2}(E^+_2e^{ik_2x} -
E^-_2e^{-ik_2x})e^{-i\omega t}$$

and finally the wave transmitted into medium (3) is 

$$E_t = E_3e^{ik_3x - i\omega t}, \,\,\,\,\, H_i =
{k_3\over \omega\mu_1}E_t$$

From these equations we see one possible problem in
using a filter rather than a single electron to interact
with an $SU(2)_L$ field: there may be numerous
reflections at the boundaries between the three media,
and these many reflections may cause a pure
$SU(2)_L$ field to be converted into an EM field,
through interacts between left and right handed
electrons in the media themselves.

\medskip
I shall assume that this does not occur.  Stratton
points out that the boundary equations are easier
manipulate in terms of the following quantities:

$$E_j = \pm Z_jH_j , \,\,\,\,\, Z_j \equiv {{\omega
\mu_j}\over k_j} , \,\,\,\,\, Z_{jk} \equiv {Z_j\over
Z_k} = {{\mu_j k_k}\over {\mu_k k_j}}$$

The boundary conditions yield four equations between
five amplitudes:

$$E_0 + E_1 = E^+_2 + E^-_2$$

$$E_0 - E_1 = Z_{12}( E^+_2 - E^-_2)$$

$$E^+_2e^{ik_2d} + E^-_2e^{-ik_2d} = E_3e^{ik_3}d$$

$$E^+_2e^{ik_2d} - E^-_2e^{-ik_2d} =
Z_{23}E_3e^{ik_3}d$$

The transmission coefficent is $ T = |E_3/E_0|^2$, so it
is only necessary to solve for

$$T = {E_3\over E_0} = {{4e^{-ik_3d}}\over
{(1-Z_{12})(1- Z_{23})e^{ik_2d} +
(1+Z_{12})(1+Z_{23})e^{-k_2d}}}$$

I shall simply by setting $\mu_1 = \mu_2 = \mu_3 =
\mu_0$, where $\mu_0$ is the magnetic permeability
of free space, and assume that $\epsilon_1 =
\epsilon_3 = \epsilon_0$, where $\epsilon_0$ is the
dielectric constant of free space.  We have in this case

$$ k_1 = k_3 = {\omega\over c}$$

\noindent
where $c$ is the speed of light in a vacuum, and

$$k_2 = \alpha + i\beta$$

\noindent
where 

$$\alpha =
{\omega\over c}\left[\left({\mu_2\over\mu_0}\right)
\left({\epsilon_2\over\epsilon_0}\right)\sqrt{1+
\left({{\sigma}\over{\epsilon_2\omega}}\right)^2} +
1\right]^{1/2}$$

\noindent
and

$$\beta =
{\omega\over c}\left[\left({\mu_2\over\mu_0}\right)
\left({\epsilon_2\over\epsilon_0}\right)\sqrt{1+
\left({{\sigma}\over{\epsilon_2\omega}}\right)^2} -
1\right]^{1/2}$$

\noindent
where $\sigma$ is the conductivity of medium (2). 
(The formulae for $\alpha$ and $\beta$ simplify in the
cases of good conductors and bad conductors --- see
([1], pp. 297 --- but with Mathematica, it's just as easy
to use the general formulae).

\medskip
the electrical conductivity is given by

$$\sigma = {ne^2\tau\over m_e}$$

\noindent
where $n$ is the number of conduction electrons per
unit volume, $e$ is the electron charge, $\tau$ is the
relaxation time, and $m_e$ is the electron mass.  This
formula for the conductivity will be valid unless the
EM frequency is greater than $5\times 10^4$ GHz.  The
key fact I shall use is that as described above,

$$n_{SU(2)_L} = {1\over2}n_{EM}$$

\noindent
since an $SU(2)_L$ field would interact with only half
of the conduction electrons.

\medskip
For conductors and semi-conductors, almost all of an
incident EM (and $SU(2)_L$) wave will be reflected
unless the thickness $d$ of the filter (medium (2)) is
small relative to the skin depth.

\bigskip
The range of wavelengths for viewing the CMBR at sea
level is 3 cm to 10 cm, or 3 GHz to 10 GHz; the upper
frequency is determined by absorption from water
vapor, and the lower frequency by background sources
in the Galaxy.  The skin depth of Copper is 1.2 microns
at 3 GHz, $\mu_{Cu}/\mu_0 = \epsilon_{Cu}/
\epsilon_0 = 1$, and $\sigma = 5.80\times 10^7\, {\rm
mho/meter}$, for which the transmission coefficient
is only $T = 0.0013$ even if $d = 10^{-3}\,({\rm skin\,
depth})$, or 12 angstroms --- the size of the copper
atom.  Clearly, no good conductor with a
conductivity comparable in magnitude to copper would
yield a detectable signal.  We want to find a material
with a transmission coefficient greater than a few per
cent for a thickness of at least 100 atoms, so that we
can trust the continuum approximation for medium (2).

\medskip
Graphite, with $\mu_{Cu}/\mu_0 = \epsilon_{Cu}/
\epsilon_0 = 1$, and $\sigma = 1.0\times 10^5\, {\rm
mho/meter}$, is marginal.  With a thickness of $d =
10^{-3}\,({\rm skin\, depth})$, or 290 angstroms ---
near the 100 atom thickness --- the transmission
coefficient is $T = 0.23$, while for $d =
(1/200)({\rm skin depth})$, or 1450 angstroms, the
transmission coefficient is $T = 0.024$, which may be
detectable.  The idea would be to allow the CMBR to
pass through the filter, and with this latter thickness,
a graphite filter would transmit 2.4\% of the CMBR
flux.  The test would be to measure the CMBR and a
reference 2.726 K reference EM source with the 
filter, using the above formulae for the relative
conductivities and for$\alpha$ and $\beta$.  More flux
will be detected from the CMBR if the pure $SU(2)_L$
hypothesis is correct.

\medskip
I shall illustrate this diffference with the material I
think has the best chance of giving an acceptable
filter, namely the element Germanium, for which
$\sigma = 2.1\, {\rm mho/meter}$ --- this
conductivity is the same at 9.2 GHz as at zero
frequency --- and $\mu_{Ge}/\mu_0 = 1$. 
Unfortunately, we have $\epsilon_{Ge}/\epsilon_0 =
16.2$ at 300 K (the ratio is 16.0 at 4.2 K), and this
non-vacuum polarizability needs special
consideration.  According to the simple models of
molecular polarizability given in ([1]. pp. 152--158),
the small temperature difference in $\epsilon_{GE}$
indicates that most of the molecular polarizability is
due to a distortion in the charge distribution by the
applied electric field generating an induced dipole
moment in each molecule.  Since the electric field will
act on {\it bound} electrons rather than {\it free}
conduction electrons, the distortion on the left handed
electrons will be almost instantaneously transmitted
to the right handed electrons, with the result that the
molecular polarizability would be the same for both an
EM field and for a pure $SU(2)_L$ field, and thus the
dielectric constants would be the same in both cases;
on the conductivities would be different, and the ratio
of the two will be 1/2, as described above.  (Even if
the effective dielectric constant were to be different
for an EM and a pure $SU(2)_L$ field, the discussion on
page 154 of [1] makes it clear that it would vary in a
more complicated way than the conductivities, and
thus the transmission coefficients would be
measurably different for the two types of gauge field. 
For example, equation 4.70 of [1] yields
$N\gamma_{mol} = 1/5$ for Germanium in an EM field;
if $N\gamma_{mol} = 1/10$ in an $SU(2)_L$ field,
equation 4.70 yields $\epsilon/epsilon_0 = 3.16$, a
substantial change from 16.2.)

\medskip
For a thickness of 1.6 millimeters, we have 

$$T_{EM}(3\, {\rm GHz}) = 0.106$$

$$T_{SU(2)_L}(3\, {\rm GHz}) = 0.132$$

which gives 

$${T_{SU(2)_L}\over T_{EM}}(3\, {\rm GHz}) = 1.24$$

\noindent
or a 24\% greater flux if the CMBR is a pure $SU(2)_L$
field.  

\medskip
The corresponding transmission coefficients at 10 Ghz
is 

$$T_{EM}(10\, {\rm GHz}) = 0.176$$

$$T_{SU(2)_L}(10\, {\rm GHz}) = 0.296$$

which gives 

$${T_{SU(2)_L}\over T_{EM}}(10\, {\rm GHz}) = 1.68$$

\noindent
or a 68\% greater flux if the CMBR is a pure $SU(2)_L$
field.  (The fluxes are greater at a higher frequency
than at the lower frequency --- opposite to what we
would expect for a good conductor --- because
Germanium is a semi-conductor.)

\medskip
A typical CMBR radiometer is pictured in Figure 8.3

\bigskip
\centerline{Figure 8.3: Figure 4.14 of Partridge, p. 126}
\bigskip

The central sky horn could be covered with the filter
--- mounting the filter inside the radiometer would
probably work, but there would be the risk of $U(1)_R$
field generation by second order effects in the wave
guide.  But Germanium is expensive, so I would risk
setting the filter inside a coaxial wave guide.

\medskip
It is important to note that the above calculations
explain why the effect I am predicting have never
been seen before.  If the parts of the radiometers
designed to absorb CMBR are thicker than the skin
depth for the radiation absorbers --- and indeed all
such absorbers are much thicker --- then all the CBMR
would be absorbed even though the effective conduction
electron density is only 1/2 of the conduction electron
density seen by the usual EM field.  In fact, the CMBR
absorbers are made much thicker than the skin depth
precisely to insure that all incident radiation will be
absorbed, and this thickness hides the effect I am
predicting.  In 1938, the German physicist G. von Droste
bombarded uranium with neutrons, but carefully
covered the uranium with metal foil in order to
eliminate alpha particles which were expected to
occur.  It worked; but the foil also eliminated fission
fragments, which have a shorter range in metal foils
that alpha particles.  In the opinion of historians ([4], p.
41) the metal foils cost von Droste the Nobel Prize for
the discovery of nuclear fission.  The same
experimental technique also cost ([8], pp. 7--8)  Enrico
Fermi a Nobel for the discovery of nuclear fission. 
Fermi began the bombardment of uranium with neutrons
in 1935, but like Droste he covered his samples with
aluminum foil.  Once again the foil absorbed the fission
fragments that in the absence of the foil, Fermi's
instruments would have clearly seen.  In the end, 
fission was discovered in 1939 by Otto Hahn and Lise
Meitner, who used not standard particle detectors, but
instead the methods of radiochemistry.  All
investigations of the CMBR to date have used too thick
a ``foil'' and thus have missed the important effect I am
predicting.  Actually, as we shall be in Section 9, there
are measurements of the CMBR that are analogous to
radiochemistry in that the instruments used do not
``cover up'' the effect: these ``instruments'' are
ultrahigh energy cosmic rays, and I shall argue that
they have already detected the effect I am predicting.

\medskip
Actually, it is possible that some early experiments
detected the expected difference between an EM CMBR
and an $SU(2)_L$ CMBR.  Two groups, one headed by
Gish and the other by Woody and Richards, measured
the CMBR using filters analogous to the filters I have
discussed above, and they detected ([3], p. 142) an
excess flux above what was expected for a 2.7 K
blackbody EM field.  The Woody and Richards experiment
also saw a lower flux than a 2.7 at lower frequencies
(below the 2.7 blackbody peak), which is just what we
would expect from an $SU(2)_L$ CMBR field, as I
discussed above.

\bigskip
\bigskip
{\bf d. Other Means of Detecting an SU(2) CMBR
Component } 
\bigskip
\bigskip
The Penning Trap is not the only way of observing an
interaction between the CMBR and a single electron.  A
Rydberg atom ---which are atoms in states with a
high principal quantum number $n$ --- can undergo
transitions induced by blackbody radiation at liquid
helium temperatures ([12], chapter 5), and hence such
atoms could be used to detect the effect I am
predicting, provided the Rydberg atom can fix its
transition electron in a definite spin state when the
atom's motion is observed.

\medskip
It has been occasionally suggested (e.g., [14]) that the
devices which allowed the observation of Bose-Einstein
condensation --- the magneto-optical trap (MOT)
--- might be able to observe the CMBR.  A MOT is
designed to excite hyperfine transitions, and the
collective motion of the atoms in the trap is
observable, so in principle the effect I am predicting
{\it might} be observable with a MOT.  The problem
with using a MOT is that the cross-section for the
hyperfine transitions is so low that MOTs usually are
set up at room temperature, and at 300 K, the tail of
the blackbody radiation in the 3 to 10 GHz range is some
two orders of magnitude greater than the 2.7 K
distribution.  The fact that low temperature
experiments can be carried out in MOTs at room
temperature is the reason MOTs are so useful.  But
this very advantage of MOTs for many experiments
makes the MOT useless for observing the CMBR.  The
opinion of Cornell and Wieman ([13], p. 49) is that ``
$\ldots$ it is difficult to imagine that [blackbody
radiation] will ever be important for trapped atoms.''

\bigskip
\centerline{\bf References}
\bigskip

\noindent\hangindent=20pt\hangafter=1
\item{[1]}J. D. Jackson, {\it Classical
Electrodynamics,Second Edition} (Wiley, New York,
1975).

\medskip
\noindent\hangindent=20pt\hangafter=1
\item{[2]} Dale R. Corson and Paul Lorrain, {\it
Introduction to Electromagentic Fields and Waves} (
Freeman, San Francisco, 1962).

\medskip
\noindent\hangindent=20pt\hangafter=1
\item{[3]} R. Bruce Partridge, {\it 3K: The Cosmic
Microwave Background Radiation} (Cambridge
University Press, Cambridge, 1995).

\medskip
\noindent\hangindent=20pt\hangafter=1
\item{[4]} Hans G. Graetzer and David L Anderson, {\it
The Discovery of Nuclear Fission} (Van Nostrand, New
York, 1962).

\medskip
\noindent\hangindent=20pt\hangafter=1
\item{[5]} Hans Dehmelt, ``Experiments on the
Structure of an Individual Elementary Particle,''
Science {\bf 247} (1990), 539--545.

\medskip
\noindent\hangindent=20pt\hangafter=1
\item{[6]} Hans Dehmelt, ``Less is More: Experiments
wiht an Individual Atomic Particle at Rest in Free
Space,'' Am. J. Phys. {\bf 58} (1990), 17--27.

\medskip
\noindent\hangindent=20pt\hangafter=1
\item{[7]} Lowell S. Brown and Gerald Gabrielse,
``Geonium Theory: Physics of a Single Electron or Ion in
a Penning Trap,'' Rev. Mod Phys. {\bf 58} (1986),
233--311.
 
\medskip
\noindent\hangindent=20pt\hangafter=1
\item{[8]} Jeremy Bernstein, {\it Hitler's Uranium
Club: The Secret Recordings at Farm Hall, Second
Edition} (Springer Verlag, New York, 2001).
 
\medskip
\noindent\hangindent=20pt\hangafter=1
\item{[9]} Hans Dehmelt, ``A Single Atomic Particle
Forever Floating at Rest in Free Space: A New Value
for Electron Radius,'' Physica Scripta {\bf T22} (1988),
102--110.
 
\medskip
\noindent\hangindent=20pt\hangafter=1
\item{[10]} P.E. Boynton, R.A. Stokes and David T.
Wilsinson, ``Primeval Fireball Intensity at $\lambda =
3.3\,{\rm mm}$,'' Phys. Rev. Lett. {\bf 21} (1968),
462--465.
 
\medskip
\noindent\hangindent=20pt\hangafter=1
\item{[11]} P.G. Roll and David T. Wilkinson, ``Cosmic
Background Radiation at 3.2 cm --- Support for Cosmic
Blackbody Radiation,'' Phys. Rev. Lett. {\bf 16},
405--407.
 
\medskip
\noindent\hangindent=20pt\hangafter=1
\item{[12]} Thomas F. Gallagher, {\it Rydberg Atoms},
(Cambridge University Press, Cambirdge, 1994).
 
\medskip
\noindent\hangindent=20pt\hangafter=1
\item{[13]} E.A. Cornell, J.R. Ensher, and C.E. Wieman,
``Experiments in Dilute Atomic Bose-Einstein
Condensation,'' in {\it Bose-Einstein Condensation in
Atomic Gases}, edited by M. Ingusio, S. Stringari,
and C.M. Wieman, (IOS Press, Amsterdam, 1999).
 
\medskip
\noindent\hangindent=20pt\hangafter=1
\item{[14]} Renato Ejnisman and Nicholas P. Bigelow,
``Is it Possible to do Experimental Cosmology Using
Cold Atoms?'' Brazilian J. of Physics {\bf 28} (1998),
72--76.
 
\medskip
\noindent\hangindent=20pt\hangafter=1
\item{[15]} Julius Adams Stratton, {\it
Electromagnetic Theory}, (McGraw-Hill, New York,
1941).

\vfill\eject

\font\bigtenrm=cmr10 scaled\magstep5
\centerline{\bf{\bigtenrm 9. Has an SU(2) CBR Component}}
\bigskip
\centerline{\bf{\bigtenrm Already Been Detected? }}
 \bigskip
\centerline{\bf{\bigtenrm Ultrahigh Energy Cosmic Rays}}

\bigskip
\bigskip

\bigskip
{\bf a. Why the Ultrahigh Energy Cosmic Rays Should Not
Exist, }   \settabs 9 \columns
\+&{\bf But Yet They Do Exist}\cr

\bigskip
\bigskip
In regard to ultra high energy (UHE) cosmic rays ---
particles above $10^{19}$ eV --- Alan Watson of the
University of Leeds recently described the
observations succinctly: ``They $\ldots$ are extremely
hard to understand: put simply --- they should not be
there''  [2].   The reason UHE cosmic rays should not be
there is that they are too energetic to be confined to
the galaxy by the galactic magnetic field, yet they
cannot propagate very far in intergalactic space
because their energy would be absorbed by collisions
with CMBR photons.

\medskip
The detailed mechanism preventing the existence of UHE
cosmic rays was discovered by Kenneth Greisen in
1966, shortly after the discovery of the CMBR. 
Greisen pointed out that protons of sufficiently high
energy would interact with the CMBR, producing pion,
resulting in a cut-off to the UHE cosmic ray spectrum. 
Even in his paper of 35 years ago, he pointed out that
``$\cdots$ even the one event recorded at $10^{20}$ eV
appears surprising. $\cdots$ [the CMBR] makes the
observed flattening ofthe primary spectrum in the range
$10^{18}-- 10^{20}$ eV quite remarkable.'' ([9], p. 750). 
Since Greisen wrote his paper, the experiments have
become even more inconsistent with the  existence of
the CMBR, as illustrated in Figure 9.1:

\bigskip
\centerline{Figure 9.1}
\medskip
Figure taken from Figure 7.1, page 119 of [1]. 
\bigskip

Figure 9.1 Caption: UHE Energy Spectrum observed
over the past 7 years with the the AGASA detector in
Japan.  The dashed curve is the expected rate with a
uniform cosmological distribution, but with the
expected interaction of protons and the CMBR.  (taken
from [1], p. 119 and [2], p. 819; figure due to M. Takeda
of the Institute for Cosmic Ray Research, University of
Tokyo [12].)  The upper group of 3 events is 6 $\sigma$
above the theretical curve.

\medskip
The AGASA array in Japan has detected 461 events
with energies above $10^{19}$ eV, and 7 events above
$10^{20}$ eV.  The Haverah Park array in England has
detected 4 events above $10^{20}$ eV, the Yakutsk
array in Siberia has detected 1 event above
$10^{20}$ eV.  The Volcano Ranch array in New Mexico
has detected 1 event above $10^{20}$ eV ([1], p.118).  So
four different groups over the past 35 years have
repeatedly seen these particles that shouldn't be there.
The consensus of the experimental cosmic ray
physicists is that the Greisen cut-off does not exist
([2], p. 818).

\medskip
At energies above $10^{20}$ eV, there is no
clustering in arrival directions ([1], p.121; [2], p.
819).  This is illustrated in Figure 9.3, which gives the
arrival directions of 114 events at energies above $4
\times10^{19}$.   At such energies, the gyromagnetic
radius is comparable to the size of the Galaxy, so UHE
cosmic rays should be extragalactic.  The only obvious
source within 30 Mpc (see below) is the Virgo Cluster,
but there is no clustering in this direction.
(Intergalactic magnetic fields are too low to effect the
arrival direction within 30 Mpc.)

\bigskip
\centerline{Figure 9.2}
\medskip
Figure taken from Figure 2, page 820 of [2]. 
\bigskip

\medskip
This blatant inconsistency between the observed
energies of some UHE cosmic rays and the global
existence of the CMBR has lead a number of physicists
to propose modifications in the known physical laws. 
However most physicists, myself included, ``$\cdots$
believe that, within certain well-defined physical
limitatons, the laws of physics as we know then can be
trusted,'' to quote the words of Malcolm Longair, the
Astronomer Royal of Scotland ([6], p. 333).  What I shall
now show is that there is a mechanism, using only the
firmly tested physical laws, whereby UHE protons can
propagate cosmological distances through the CMBR
--- provided the CMBR is not the complete EM field,
but rather only the $SU(2)_L$ part of the EM field.

\bigskip
{\bf b. How an SU(2) Componet to the CBR Would Permit}
\+& {\bf UHE Cosmic Rays to Propagate}\cr

\bigskip
Recall that CMBR blocks the propagation of UHE cosmic
rays via the GZK effect [9]: protons comprising the UHE
particles collide with a CMBR photon, resulting in pion
production.  The reactions are

$$\gamma + p \rightarrow \Delta^+ \rightarrow \pi^0
+ p\eqno(9.1)$$

$$\gamma + p \rightarrow \Delta^+ \rightarrow \pi^+
+ n\eqno(9.2)$$

$$\gamma + p \rightarrow \Delta^{++} + \pi^-
\rightarrow \pi^- + \pi^+  + p\eqno(9.3)$$

\noindent
where $p$, $n$, and $\pi$ are the proton, neutron, and
pion respectively.  The reaction cross-sections are
dominated by $\Delta$ particle resonances ([5], [9]). 
Of the total cross-section for $(9.2)$ of 300
microbarns at peak $E_\gamma = 320$ MeV, 270
comes from the $\Delta$ resonance ([5], p. 14).  Of the
total cross-section for $(9.3)$ of 70 microbarns at
peak $E_\gamma = 640$ MeV, virtually all comes from
the $\Delta^{++}$ resonance ([5], p. 14).  Of the total
cross-section for $(9.1)$ of 250 microbarns at peak
$E_\gamma = 320$ MeV, 140 comes from the $\Delta$
resonance ([5], p. 13).  For $(9.3)$, virtually all the
total cross-section for photon energies less than the
peak also comes from the $\Delta^{++}$ resonance.
For the other reactions, the rest of the total
cross-section arises from a photoelectric term ([5], p.
14).					

\medskip
However, if the CMBR consists not of electromagentic
radiation, but is instead a pure $SU(2)_L$ field, then
the $\Delta$ resonance cannot occur.  The reason is
simple: the $\Delta$ particle originates from a proton
by a quark spin flip ([5], p. 16), but since a $SU(2)_L$
field couples only to a left-handed quark, it cannot
induce a quark spin flip: a pure $SU(2)_L$ photon would
not couple to a right-handed quark at all, and a
left-handed quark would have the handedness
unchanged by the interaction.

\medskip
Furthermore, the photoelectric term would be reduced
because only a fraction of the electric charge on the
quarks would interact with a pure $SU(2)_L$ field.  If
for example, a proton were to have only its down
valence quark left handed, then its effective electric
charge would be $-1/3$ rather than $+1$.  Since the
photo cross-sections are proportional to the
$({\rm charge} )^4$ (the square of the classical
electron radius, with the ``electron'' having a charge of
-1/3), the photo cross-section would be reduced by a
factor of $1/81$ from its value for an electromagentic
CMBR.  Even it one of the up quarks were left-handed,
the photo cross-section would be reduced by a factor
of $16/81 \approx 1/5$.

\medskip
The net effect on the total-cross-sections (for the
down valence quark case) is to reduce the
cross-section for pion production from $SU(2)_L$
photons $\sigma_{SU(2)}$ from its value
$\sigma_{EM}$ that we would have if the CMBR were
to be a normal electomagnetic field:

$$\sigma_{SU(2)}^{p\pi^0} =
{1\over150}\sigma_{EM}^{p\pi^0}\eqno(9.4)$$

$$\sigma_{SU(2)}^{n\pi^+} =
{1\over810}\sigma_{EM}^{n\pi^+}\eqno(9.5)$$

$$\sigma_{SU(2)}^{p\pi^+\pi^-} = 0\eqno(9.6)$$

The mean free path $L_{MFP}$for an UHE proton is

$$L_{MFP} = (\sigma^{p\pi^0}N_{photon})^{-1}$$

Using $N_{photon} = 5\times10^8\,{\rm m}^{-3}$ we
would get $L_{MFP|} \approx 10^{23}\,{\rm m} \approx
3\, {\rm Mpc}$ ([6], p. 340) if we used
$\sigma_{EM}^{p\pi^0}$. Using $(9.4)$, however, we get

$$L_{MFP} = 450\,{\rm Mpc} \eqno(9.7)$$

\noindent
which means that UHE protons can propagate through
the intergalactic medium as if the CMBR were not
there.  This becomes even more obvious when we
consider that the fractional energy loss due to pion
creation is $\Delta E/E \approx m_\pi/m_p \approx
{1\over10}$, so the propagation distance would be
more than 4 Gpc, which is truly a cosmological
distance.

\medskip
If pion production is no longer significant, then one
could wonder about the removal of UHE proton via
electron-positron pair production.  As is well-known
([6], p. 341), the cross-section for pair production
from the collision of a UHE proton with an EM CMBR
photon is actually greater than the cross-section for
pion production, but the much smaller mass of the pair
means that with EM CMBR photons, the energy loss
per collision is less by a factor of
$(m_\pi/m_e)(\sigma^{p\pi}/\sigma_{pair}) \approx
6$. I have shown in an earlier section of this paper that
pair production is not possible via collision of two
$SU(2)_L$ photons, but it is not necessary to
investigate whether this would be the case for the
collision of such a CMBR photon with a proton.  For the
cross-section for pair production is proportional to
$\alpha_{EM}r^2_e$, and thus the cross-section for
pair production would be also reduced by at least a
factor of $1\over150$ by the effective charge
reduction mechanism that reduced the pion production
cross-section.

\medskip
The energetics of the UHE cosmic rays are completely
consistent with a cosmological source ([6], pp.
341--343).  The local energy density of cosmic rays
with energies above
$10^{19}\,{\rm eV}$ is $1\,{\rm eV m}^{-3}$. 
Following [6], p. 342), let us assume that each source
of UHE cosmic rays generates a total cosmic ray
energy of $E_{CR}$ over the age of the universe, and
that $N$ is the spatial density of these sources.  In
the absence of losses, the energy density of UHE
cosmic rays would be

$$\rho_{CR} = E_{CR}N$$

For strong radio galaxies, $N\approx 10^{-5}\, {\rm
Mpc}^{-3}$, so we would have to have $E_{CR} \approx
5\times10^{53}\,{\rm J}$, or about $3\times
10^7\,{\rm M_\odot}$ of energy in the form of UHE
cosmic rays produced per source over the history of
the universe.  Given that the black hole cores of many
radio galaxies have masses above
$10^{10}\,{\rm M_\odot}$, one would require a
conversion efficiency of mass into UHE cosmic rays of
only 0.6\% (assuming that the source of the UHE
cosmic rays produces these protons with left-handed
and right-handed down valence quarks in equal
numbers), which seems quite reasonable: even ordinary
hydrogen fusion has a 0.7\% efficiency for conversion
of mass into energy, and black hole models can give
mass-energy conversion efficiencies up to 42\%. 

\medskip
The sources for a $3 \times 10^{20} {\rm eV}$
proton which are allowed by the Hillas criterion,
namely that an allowed source must satisfy
$BR \sim 10^{18} {\rm G\, cm}$, where $B$ is the
manetic field and $R$ is the size of the region with
this roughly constant field, are radio galaxy lobes and
AGNs, as is well-known.  However, heretofore, these
sources have been eliminated on the grounds that they
are too far away.  If the CMBR is actually a pure
$SU(2)_L$ field, then such sources are perfectly
acceptable.

\bigskip
\centerline{\bf 9.c Cosmic Ray Physicists Have Once
Again}
\+&{\bf Seen New Fundamental Physics }\cr

\bigskip
Cosmic ray physicists have in the past made great
discoveries in fundamental physics: in 1932, the
discovery of the positron ([3], reprinted in [4]); in 1937
the discovery of muons; and in 1947, the discovery of
pions and kaons ([4], pp. 50--51).  Positrons were the
first examples of anti-matter, and finding them
deservedly got Carl Anderson the Noble Prize.  Muons
are elementary particles according to the Standard
Model, as are s-quarks, which made their first
appearance in kaons.  The first s-quark baryons, the
$\Lambda$, the $\Xi^\pm$, and the $\Sigma^+$
particles, were first detected by cosmic ray
physicists, in 1951, in 1952, and in 1953, respectively
([4], pp. 50--51).   But it has been almost half a century
since cosmic ray physicists have made {\it recognized}
fundamental discoveries.  I believe that the discovery
of the UHE cosmic rays are an {\it unrecognized}
discovery of fundamental importance: the observation
of these particles demonstrates that the CBR is not an
electromagentic field, but rather the pure SU(2)
component of an electromagnetic field.  

\medskip
The expression of many theorists concerning UHE
cosmic rays, that these particles simply {\it must} be
merely a local phenomena, reminds me of Herzberg's
1950 remark on the observation that CN molecules in
interstellar clouds appeared to be in a heat bath of
around 2.3 K: ``which has of course only a very
restricted meaning.'' ([11], p. 497), by which he mean
that the 2.3 heat bath was merely a phenomena local to
the molecular cloud.

\centerline{\bf{References}}
\bigskip
\noindent\hangindent=20pt\hangafter=1
\item{[1]}  Friedlander, Michael W. 2000 {\it A Thin Cosmic Rain: Particles from Outer Space}
(Cambridge: Harvard University Press).

\medskip
\noindent\hangindent=20pt\hangafter=1
\item{[2]}  Watson, Alan 2001 ``Ultra High Energy
Cosmic Rays:  What we Know Now and What the Future
Holds," in {\it Relativistic Astrophysics: 20th Texas
Symposium (AIP Conference Procedings, volume 586},
J. C. Wheeler and H. Martel (eds.) American Institure of
Physics, pp. 817--826.  (The original abstract was
different from the published abstract)

\medskip
\noindent\hangindent=20pt\hangafter=1
\item{[3]} Anderson, Carl D. 1933. ``The Positive Electron,'' Phys. Rev. {\bf 43}, 491--494.

\medskip
\noindent\hangindent=20pt\hangafter=1
\item{[4]} Hillas, A. M. 1972. {\it Cosmic Rays} (Oxford: Pergamon Press)

\medskip
\noindent\hangindent=20pt\hangafter=1
\item{[5]}  N. Nagle, V. Devanathan, and H. \"Uberall, {\it Nuclear Pion Production}
(Berlin: Springer-Verlag, 1991).

\medskip
\noindent\hangindent=20pt\hangafter=1
\item{[6]} Malcolm S. Longair, {\it High Energy
Astrophysics, Volume 2: Stars, the Galaxy, and the
Interstellar Medium [Second Edition]} (Cambridge
University Press, Cambridge, 1994).

\medskip
\noindent\hangindent=20pt\hangafter=1
\item{[7]} James W. Cronin, ``Cosmic Rays: the Most
Energetic Particles in the Universe," Rev. Mod. Phys.
{\bf 71} (1999), S165--S172.

\medskip
\noindent\hangindent=20pt\hangafter=1
\item{[8]} Pierre Sokolsky, {\it Introduction to
Ultrahigh Energy Cosmic Ray Physics}
(Addision-Wesley, New York, 1989).

\medskip
\noindent\hangindent=20pt\hangafter=1
\item{[9]} Kenneth Greisen, ``End to the Cosmic-Ray
Spectrum?'' Phys. Rev. Lett. {\bf 16} (1966), 748--750.

\medskip
\noindent\hangindent=20pt\hangafter=1
\item{[10]} D. W. Sciama, ``The Impact of the CMB
Discovery on Theoretical Cosmology,'' in {\it The
Cosmic Microwave Background: 25 Years Later}, edited
by N. Mandolesi and N. Vittorio (Kluwer, Amsterdam,
1990).

\medskip
\noindent\hangindent=20pt\hangafter=1
\item{[11]} Gerhard Herzberg, {\it Molecular Spectra
and Molecular Sturcture: I. Spectra of Diatomic
Molecules, Second Edition} {Van Nostrand, New York,
1950).

\medskip
\noindent\hangindent=20pt\hangafter=1
\item{[12]} M. Takeda et al, Phys. Rev. Lett. {\bf 81}
(1998), 1163.

\vfill\eject
\bye